\documentclass[journal]{IEEEtran}
\usepackage{amsthm,amsmath,amssymb}
\usepackage{eucal}
\usepackage{xifthen}
\usepackage{mathtools}
\usepackage{enumerate}
\usepackage{microtype}
\usepackage{xspace}
\usepackage{bm}
\usepackage{cite}
\usepackage{fancyhdr}
\usepackage{lastpage}
\usepackage[small]{caption}
\usepackage{xcolor}
\allowdisplaybreaks

\long\def\comment#1{}

\newfont{\bbb}{msbm10 scaled 700}

\newfont{\bb}{msbm10 scaled 1100}
\newcommand{\CC}{\mbox{\bb C}}

\newcommand{\mbs}[1]{\bm{#1}}
\newcommand{\vect}[1]{{\lowercase{\mbs{#1}}}}
\newcommand{\mat}[1]{{\uppercase{\mbs{#1}}}}

\newcommand{\T}{{\scriptscriptstyle\mathsf{T}}}

\renewcommand{\Re}[1][]{\ifthenelse{\isempty{#1}}{\operatorname{Re}}{\operatorname{Re}\left(#1\right)}}
\renewcommand{\Im}[1][]{\ifthenelse{\isempty{#1}}{\operatorname{Im}}{\operatorname{Im}\left(#1\right)}}

\newcommand{\lv}{\vect{l}}

\newcommand{\pv}{\vect{p}}
\newcommand{\qv}{\vect{q}}

\newcommand{\uv}{\vect{u}}
\newcommand{\vv}{\vect{v}}
\newcommand{\wv}{\vect{w}}
\newcommand{\xv}{\vect{x}}

\newcommand{\Am}{\mat{a}}

\newcommand{\Cm}{\mat{c}}

\newcommand{\Pm}{\mat{p}}

\newcommand{\Wm}{\mat{w}}
\newcommand{\Xm}{\mat{x}}

\newcommand{\Ec}{{\mathcal E}}

\newcommand{\Gc}{{\mathcal G}}

\newcommand{\Kc}{{\mathcal K}}

\newcommand{\Mc}{{\mathcal M}}

\newcommand{\Pc}{{\mathcal P}}

\newcommand{\Xc}{{\mathcal X}}

\newcommand{\one}{\mat{\mathrm{1}}}
\newcommand{\zero}{\mat{\mathrm{0}}}

\newcommand{\CN}[1][]{\ifthenelse{\isempty{#1}}{\mathcal{N}_{\mathbb{C}}}{\mathcal{N}_{\mathbb{C}}\left(#1\right)}}
\renewcommand{\P}[1][]{\ifthenelse{\isempty{#1}}{\mathbb{P}}{\mathbb{P}\left(#1\right)}}
\newcommand{\E}[1][]{\ifthenelse{\isempty{#1}}{\mathbb{E}}{\mathbb{E}\left(#1\right)}}
\renewcommand{\det}[1][]{\ifthenelse{\isempty{#1}}{\mathrm{det}}{\mathrm{det}\left(#1\right)}}
\newcommand{\trace}[1][]{\ifthenelse{\isempty{#1}}{\mathrm{tr}}{\mathrm{tr}\left(#1\right)}}
\newcommand{\rank}[1][]{\ifthenelse{\isempty{#1}}{\mathrm{rank}}{\mathrm{rank}\left(#1\right)}}
\newcommand{\diag}[1][]{\ifthenelse{\isempty{#1}}{\mathrm{diag}}{\mathrm{diag}\left(#1\right)}}

\DeclarePairedDelimiter\norm{\lVert}{\rVert_2}
\DeclarePairedDelimiter\Norm{\lVert}{\rVert_2^2}

\renewcommand{\arg}{{\hbox{arg}}}

\renewcommand{\Re}{{\rm Re}}
\renewcommand{\Im}{{\rm Im}}

\newcommand{\herm}{{\sf H}}

\newcommand{\transp}{{\sf T}}

\allowdisplaybreaks
\newcommand{\st}{{\rm s.t.}}
\DeclareMathAlphabet{\mathcal}{OMS}{cmsy}{m}{n}

\newtheorem{remark}{Remark}

 \usepackage{hyperref}
 \hypersetup{
     bookmarks=false,         
     unicode=false,          
     pdftoolbar=true,        
     pdfmenubar=true,        
     pdffitwindow=false,     
     pdfstartview={FitH},    
     pdfnewwindow=true,      
     colorlinks=true,       
     linkcolor=red,          
     citecolor=green,        
     filecolor=magenta,      
     urlcolor=cyan           
 }
\usepackage{subfigure}
\usepackage{cite}
\usepackage[final]{graphicx}
\usepackage{amsmath}
\usepackage{amssymb}
\usepackage{amsfonts}
\usepackage{array}
\usepackage{amsfonts}
\usepackage{stfloats}
\usepackage{lineno}
\usepackage{colortbl}
\usepackage{algorithmic}
\usepackage{epstopdf}
\usepackage{amsthm}
\usepackage{multirow}
\usepackage{multicol}
\usepackage{graphicx}
\usepackage{lipsum}
\usepackage[margin=0.7in]{geometry}
\newcolumntype{C}{>{\centering\arraybackslash$}p{\linewidth}<{$}}
\usepackage[linesnumbered,ruled,vlined]{algorithm2e}
\theoremstyle{plain}
\theoremstyle{remark}
\usepackage{caption}
\usepackage{subfigure}

\allowdisplaybreaks  

\begin{document}
\title{Topology-Aware Integrated Communication, Sensing, and Power Transfer for Multi-User SAGIN}

\author{Han Yu,~\IEEEmembership{Member, IEEE,}~Jiajun He,~\IEEEmembership{Member, IEEE,}~Xinping Yi,~\IEEEmembership{Senior Member, IEEE}\\
Feng Yin,~\IEEEmembership{Senior Member, IEEE,}~Hing Cheung So,~\IEEEmembership{Fellow, IEEE,}~Giuseppe Caire,~\IEEEmembership{Fellow, IEEE}
\thanks{Parts of this paper were presented in part at the IEEE ICC 2026 \cite{yu-icc}.}
\thanks{H. Yu and G. Caire are with the Faculty of Electrical Engineering and Computer Science, Technical University of Berlin, Germany (email:han.yu.1@campus.tu-berlin.de, caire@tu-berlin.de).}
\thanks{J. He is with the Centre for Wireless Innovation (CWI), Queen’s University Belfast, BT3 9DT Belfast, U.K. (email: j.he@qub.ac.uk).}
\thanks{X. Yi is with the National Mobile Communications Research Laboratory, Southeast University, Nanjing 210096, China (e-mail: xyi@seu.edu.cn).}
\thanks{F. Yin is with the School of Artificial Intelligence (SAI), The Chinese University of Hong Kong, Shenzhen, 518172, China (e-mail: yinfeng@cuhk.edu.cn).}
\thanks{H. C. So is with the Department of Electrical Engineering, City University of Hong Kong, Hong Kong SAR, China (email: hcso@cityu.edu.hk).}
}

\maketitle

\begin{abstract}
In sixth-generation and beyond, space-air-ground integrated networks (SAGINs) extend network connectivity to space, thereby enabling broader service coverage. This paper proposes a topology-aware SAGIN framework to address the integrated sensing, communication, and wireless power transfer (ISCPT) problem, leveraging the distinctive visibility of satellite-terrestrial and satellite-satellite users as well as their constructing in-between channel strengths. By modeling the topology of the SAGIN as a bipartite graph, we formulate the ISCPT problem as a multi-objective joint optimization problem with specified topological structures to reflect connection relationships of satellite-terrestrial and satellite-satellite users. 
The ISCPT problem is then reformulated and carefully decomposed as several mixed-integer linear programs (MILPs) by leveraging the network topology to individually optimize sensing, communication, and power transfer. To reduce the computational complexity of the proposed method, a greedy algorithm deal with generalized multi-assignment problem (GMAP) is developed. Simulation results demonstrate superior performance in communication and sensing, with a tolerable trade-off in wireless power transfer.
\end{abstract}


\section{Introduction}

The next generation of communication systems shall accommodate massive concurrent mobile connections, surpassing the capacity limits of conventional terrestrial infrastructures. Space-air-ground integrated networks (SAGINs) have emerged as a promising solution to provide global, resilient, and seamless connectivity for both terrestrial and non-terrestrial users, \cite{ye2020SAGIN, liu2018SAGIN, wang2020SAGIN}. By leveraging the dense deployment of satellite access points (APs), SAGIN enables flexible trade-offs among latency, coverage, and energy efficiency. Compared to terrestrial networks \cite{Chien2024iot}, SAGIN offers significantly broader coverage in undeserved or harsh environments where terrestrial infrastructure may be unavailable or unreliable. Moreover, inter-satellite cooperation across multi-layer constellations substantially enhances overall system performance. Future satellite networks are envisioned to transcend conventional connectivity by integrating communication, sensing, and energy delivery into a unified multi-functional platform \cite{liu2022isac, he2026LAWN}. However, realizing such integration at scale presents formidable challenges, including satellite scheduling, resource allocation, and cross-layer coordination, requiring holistic and scalable algorithmic solutions.

\subsection{Motivations}

SAGIN is expected to support massive connectivity for both terrestrial and space users while concurrently enabling multiple mission-critical functionalities. In general, three representative functionalities will be considered in satellite networks, including: 

\begin{enumerate}
    \item \textit{Communication}: Satellite APs are required not only to communicate with terrestrial and remote ground users, but also to exchange information with other satellites and spaceborne users for efficient coordination \cite{SIN}. To ensure real-time and ubiquitous connectivity for a massive number of users, satellites across both intra- and inter-orbit layers are interconnected through inter-satellite links (ISLs) and inter-orbit links \cite{Yang_CSIN}.
    
    \item \textit{Positioning, Navigation, and Timing (PNT)}: PNT is a fundamental yet critical function for both terrestrial and non-terrestrial users \cite{10040628}. For terrestrial users, such as mobile devices and Internet-of-Vehicles, precise location, velocity, and timing information are essential for accurate localization, navigation, and real-time tracking. For spaceborne users, accurate spatiotemporal information is vital for efficient coordination within multi-layer satellite networks. For instance, reliable knowledge of satellite positions helps prevent collisions and supports seamless inter-satellite operations.
    
    \item \textit{Wireless Power Transfer (WPT)}: Power availability is a critical concern for satellite networks to ensure sustained and efficient operation. Typically, satellites, especially those in medium Earth orbit and geostationary Earth orbit, harvest solar energy and convert it into electrical power to support onboard functions \cite{7498851}. Beyond self-sustenance, satellites can also convert harvested solar energy into direct current power, and subsequently into radio frequency signals for WPT.
\end{enumerate}

This work investigates the integration of three representative functionalities, including sensing, communication, and WPT, within a SAGIN. Executing these tasks in parallel presents inherent challenges due to limited onboard resources, strict power constraints, and spectrum reuse limitations. Moreover, balancing resource allocation across different satellite constellations is critical to ensure reliable and efficient performance across all functionalities. Interference from both terrestrial and non-terrestrial users further complicates system coordination and shall be carefully mitigated. These challenges highlight the necessity of a scalable and intelligent management framework capable of orchestrating heterogeneous service requirements. Thus, we aim to propose an efficient strategy to maximize the overall performance of multi-user integrated sensing, communication, and wireless power transfer (ISCPT) system.

\vspace{-0.5cm}

\begin{table*}[t]
    \centering
    \footnotesize
    {{
    \caption{Nomenclature}~\label{tab:system_parameters}
    \vspace{-0.15cm}
    \resizebox{17cm}{!}
    {
    \begin{tabular}
    {p{0.8cm}|p{5cm}|p{0.8cm}|p{5cm}|p{0.8cm}|p{5cm}}
    \hline
    \hline
    Symbol & Meaning & Symbol & Meaning & Symbol & Meaning \\
    \hline
    $\Kc_{C,T}$& Set of terrestrial communication users  & $\Kc_{C,S}$ & Satellite set acting as communication users & 
    $\Kc_{\text{Sen}}$& Satellite set acting as sensing target \\
    $\Kc_P$ & Satellite set acting as charging users
    & $\Kc_S$ & Set of satellite users & $\Kc_C$ & Set of communication users \\
    $L$ & Longitude & $H$ & Height& $B$ & Latitude \\ 
    $N$ & Prime vertical radius of curvature & $w_{mk}^2$ & Path-loss strength between the $m$-th AP and $k$-th user &$\rho_m$ & Transmit power of AP $m$\\
    $\phi_{mk}$ & Elevation angle between AP $m$ and user $k$ & $\tau_\phi$ & Elevation angle threshold & $h_{mk}$ & Channel between AP $m$ and user $k$ \\
    $G_U$ & Antenna gain at the user side & $G_B$& Antenna gain at the AP side & $R_c$ & Sum rate of communication users \\
    $\gamma_k$ & SINR of communication user $k$ & $\gamma_s$ & Sensing SINR &
    $E_p$ & Received power by charging users \\
    $\Gc$ & Bipartite Graph & $\Mc$ & Set of all APs&$e(m,k)$& Edge between $m$-th AP and $k$-th user \\
    $\Kc$ &Set of all users & $\Ec$ & Edges of graph $\Gc$& $\lv_a$ & The position of user or AP under ECEF coordinates, $a\in \Kc\cup \Mc$\\
    $\tau_{\text{PL}}$ & Channel strength threshold & $\Am$ & Adjacency matrix of graph $\Gc$ & $\Wm$ & Weighting matrix \\
    $R_{c,\text{T}}$ & TA-based sum rate&$\gamma_{k,\text{T}}$ & TA-based rate of communication user $k$ & $R_{\text{Earth}}$ & Earth radius \\ $u_k$ & Communication user status, binary
    &$v_m$ & AP status, binary & $v_m^s$& Sensing target AP selection status, binary\\
    $E_{p,\text{T}}$ & TA-based received energy & $\rho_{\max}$ & Maximum transmit power & $\hat{\tau}_{c,m}$ & TA-based communication threshold at AP $m$ side \\
    $\hat{\tau}_p$ & Power transfer threshold & $\hat{\tau}_{s,k}$ & TA-based threshold of sensing interference from the $k$-th communication user
    & $\Pm_c$ & Profit matrix of communication users\\
    $ \pv_s$ & Profit vector of the sensing target & $\Pm_p$ & Profit matrix of charging users & $\Cm_c$ & Cost matrix of communication users\\
    $\Cm_s$ & Cost matrix of the sensing target & $\qv^c$ & Final profit vector of communication users & $q^s$ & Final profit of the sensing target \\
    $q^p$ & Final profit of power transfer & $ N_\tau$ & Threshold of serving communication user & $\tau_s^g$ & Sensing threshold of Greedy algorithm \\
    \hline 
    \end{tabular}}}}
\end{table*}

\subsection{Related Works}

Integrated sensing and communication (ISAC) has been extensively investigated in terrestrial networks, where ISAC APs are designed to provide both high-rate connectivity and high-precision environmental sensing within a unified infrastructure \cite{liu2022isac}. Early research primarily focused on dual-functional APs serving multiple communication users while detecting a single target \cite{liu2022single}. By further exploiting information embedded in reflected signals, target localization and tracking can be achieved using time-delay and angle-of-arrival (AoA) measurements \cite{win2023loc, liu2022track, he-gc}. Beyond enhanced sensing, recent studies have explored integrating ISAC with other key functionalities. In \cite{he2024iscc}, an integrated sensing, communication, and computation (ISCC) framework is proposed, aiming to maximize sensing performance under a given delay constraint. The problem is formulated as a convex optimization task, where the sensing threshold and sampling rate are jointly optimized. To reduce computation latency, \cite{yao2025iscc} introduces federated edge learning in unmanned aerial vehicle (UAV)-assisted ISCC systems, where a joint UAV deployment and ISCC resource allocation problem is formulated to minimize the total training time. Given the mixed-integer and non-convex nature of the problem, an alternating optimization approach is adopted, which decomposes the original problem into three subproblems to obtain sub-optimal solutions. In parallel, WPT is gaining attention for integration due to its critical role in supporting energy-limited operations. For example, \cite{li2024iscpt} suggests an ISCPT architecture that unifies ISAC and simultaneous wireless information and power transfer (SWIPT) to increase both energy and spectral efficiency. A unified analytical framework for ISCPT is further developed in \cite{zhou2024iscpt}, where the joint optimization of sensing, communication, and power transfer is formulated as a non-convex problem and efficiently solved using semidefinite relaxation (SDR) and Schur complement transformation.

While such integration has been well studied in terrestrial networks, extending multi-functional ISAC to SAGIN remains an open research challenge. In \cite{yin2024leoisac}, the feasibility of ISAC in low Earth orbit (LEO) satellite networks is explored, highlighting rate-splitting multiple access (RSMA) as a promising technique for balancing sensing and communication. A deep learning-based predictive optimization network is designed in \cite{lukito2025leoisac} to jointly optimize the transmit waveform and receive filter, thereby eliminating explicit channel estimation and reducing signaling overhead. Liu \textit{et al.} \cite{liu2024rsma} investigate an RSMA-assisted LEO ISAC system, where the non-convex problem of maximizing the minimum energy efficiency under joint sensing and communication constraints is reformulated as a semidefinite programming (SDP) problem. An iterative algorithm combining successive convex approximation (SCA) and Dinkelbach’s method is proposed to derive the optimal precoding solutions. In addition, \cite{pala2025leoisac} develops an analytical framework for reconfigurable intelligent surface-assisted satellite–terrestrial integrated networks, leveraging a federated deep deterministic policy gradient algorithm to improve spectral efficiency through distributed resource optimization.

{Recent studies have investigated multi-satellite networks in different application scenarios \cite{zhang2026gnn,ni2025llm,zhang2026secure}. Specifically, Zhang $et$ $al.$ \cite{zhang2026gnn} developed a graph neural network-based framework for physical layer security in multi-satellite collaborative systems, where satellite-user association and beamformer are jointly designed to improve the secrecy rate. In \cite{ni2025llm}, a large language model-aided spectrum-sharing framework was proposed for LEO satellite communications to support dynamic resource allocation and service-aware transmission in highly time-varying environments. Zhang $et$ $al.$ \cite{zhang2026secure} further integrated ISAC into a multi-satellite system and investigated the joint design of satellite assignment and beamformer, thereby extending the problem from communication-only security to a secure ISAC framework. Nevertheless, in multi-satellite networks, conventional optimization methods typically incur high computational complexity and often require complete redesign when the optimization objectives vary, which limits their practicality in dynamic multi-satellite environments. 
In addition, although these methods demonstrate promising performance, the underlying physical insights are often obscured by complex optimization formulations or data-driven architectures, thereby limiting the interpretability of the resulting designs. Consequently, it calls for a more unified, and interpretable framework for multi-satellite coordination networks.} 
\subsection{Contributions}

To address the aforementioned limitations, we devise a topology-aware (TA) framework that leverages the structural characteristics of multi-satellite, multi-user SAGINs to jointly optimize communication, sensing, and WPT performance. Compared to terrestrial networks, multi-satellite networks exhibit relatively stable topologies, as their connectivity is primarily influenced by large-scale fading factors such as distance and coverage. In contrast to existing methods, the proposed topology-based approach provides clear physical interpretations of the objectives within the optimization constraints, offering improved generalizability, enhanced flexibility, and greater practical applicability. The main contributions of this work are summarized as follows:

\begin{itemize}
\item \textit{Topology-Aware Satellite Framework:}
We propose a unified TA framework for joint terrestrial and non-terrestrial multi-user SAGINs by modeling visibility and channel strength through a sparse bipartite graph. This graph effectively captures the dominant connectivity between satellite APs and users. Compared to conventional approaches that address individual objectives separately, our framework supports both single- and multi-objective formulations within a unified structure. Moreover, the inherent sparsity of the graph reduces algorithmic complexity, thereby improving scalability.

\item \textit{Multi-Objective Problem Formulation:}
Within the proposed TA framework, we formulate a multi-objective optimization problem that jointly considers communication, sensing, and WPT. The objectives, sum rate, sensing signal-to-interference-plus-noise ratio (SINR), and transferred power, are embedded into the graph structure via connectivity and edge weights. This establishment offers a flexible foundation for incorporating additional objectives and adapting to diverse application scenarios.

\item \textit{Topology-Assisted Algorithm Design:}
To solve the problem efficiently, we reformulate and decompose the ISCPT problem as a couple of mixed-integer linear programs (MILPs) by introducing relevant threshold constraints. An alternating optimization strategy is employed to identify suitable hyperparameters and ensure convergence. In addition, a low-complexity greedy algorithm, inspired by the generalized multi-assignment problem (GMAP), is developed to achieve near-optimal performance with significantly reduced computational overhead. Extensive simulations demonstrate that the developed approach significantly outperforms state-of-the-art (SoTA) benchmarks across both communication and sensing metrics under a tolerable reduction of power transfer performance.
\end{itemize}

This paper is structured as follows. Section \ref{sec:SecII} illustrates the ISCPT problem within the specified distributed SAGIN. Section \ref{sec:topo} discusses the topological structure through applying two criteria, namely, visibility and channel strength, and the re-presentation of three objectives within this framework. Section \ref{sec:OptPro} defines TA multi-objective optimization. Section \ref{sec:simulation} reports the simulation results that evaluate the performance of our TA-based method, accompanied by a comprehensive discussion. Section \ref{sec:con} provides the conclusion of this paper. 

{\textit{Notation:}} In this paper, we denote the scalar, vector, matrix, and set by $x, \xv, \Xm, \Xc$, respectively. Let $\xv^\transp$ and $\xv^\herm$ represent the transpose and conjugate transpose of the vector $\xv$. In addition, $[\xv]_i$ denotes the $i$-th element of $\xv$, and $[\Xm]_{i,j}$ refers to the element located in the $i$-th row and $j$-th column of the matrix $\Xm$. The symbol $\j$ denotes the imaginary unit, while $|\Xc|$ represents the cardinality of the set. The notation $\norm{\xv}$ represents the $l_2$-norm of $\xv$. We define $\one_M$ and $\zero_M$ as the $M$-dimensional one vector and zero vector, respectively. {The key notations used in the paper are summarized in Table \ref{tab:system_parameters}.}

\section{System Model}
\label{sec:SecII}
\begin{figure}
    \centering
    \includegraphics[width=\columnwidth]{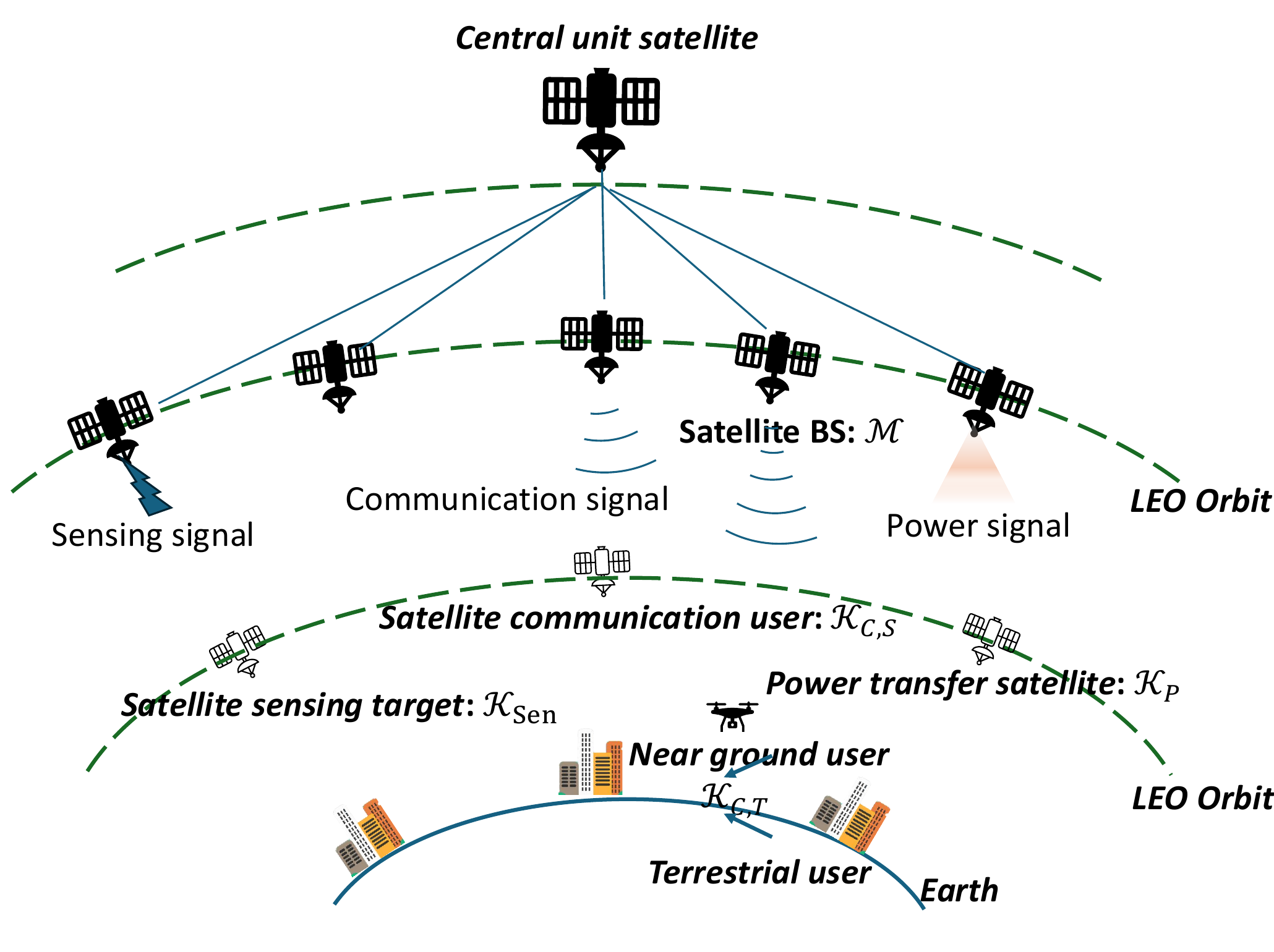}
    \caption{SAGIN comprises APs and multiple terrestrial and satellite users. This network consists of the LEO constellation in higher orbits, which includes multiple satellite APs. A subset of LEO satellites acts as satellite users, facilitating communication, sensing, and wireless power transfer.
 In addition, the network incorporates terrestrial users for communication purposes. A central unit (CU) satellite exists that controls the satellite APs. The CU has only the topology structure and provides the strategy to satellite APs.}
    \label{fig:SysModel}
\vspace{-0.5cm}
\end{figure}

As illustrated in Fig. \ref{fig:SysModel}, we consider a multi-user SAGIN comprising $M$ LEO satellites that function as APs, along with multiple satellite users supporting communication, sensing, and WPT functionalities. Terrestrial users also request communication services from the satellites. {A central unit (CU) satellite is considered to facilitate the optimization and decision-making processes. The CU only needs knowledge of the topological structure; therefore, information exchange between the CU and the APs is only required occasionally. Once the topological structure changes, the CU reassigns the corresponding strategy to the APs. Given the highly dynamic yet predictable operating environment, such strategy updates are feasible. Nevertheless, the update problem under dynamic SAGINs remains a significant challenge in the proposed TA-based framework and will be investigated in future work.} For clarity, we define the following user sets: $\Kc_{C, T}$ denotes the set of terrestrial communication users, while $\Kc_{C, S}$ represents the set of satellites acting as communication users. Furthermore, within the remaining satellite users, two sets are defined: 1) $\Kc_P$, the set of satellites acting as charging users enabled by WPT, and 2) $\Kc_{\text{Sen}}$, the set of satellites acting as the sensing target, where $|\Kc_\text{Sen}|=1$. In addition, our system defines the term `terrestrial user' to encompass both terrestrial and near-ground users, e.g., UAVs.

{
\begin{remark}
    The single-target assumption has been widely adopted in the existing ISAC literature \cite{su2021security, wang2026security, wei2024UAV}. For instance, \cite{su2021security} considers a scenario in which an eavesdropper (Eve) intercepts information transmitted by a nearby base station (BS), and part of the system resources is allocated to detect or localize Eve. In addition, most existing studies on high-precision localization and environmental mapping focus on a single target, whilst multi-target sensing introduces additional challenges, including measurement association and increased resource consumption \cite{yu2025icc, 10271553}, which require further investigation in the future work.
\end{remark}
}

\vspace{-0.5cm}

\subsection{Coordinate Transformation}

Before establishing the objectives for communication, sensing, and WPT, we first introduce the coordinate transformation for the satellite APs and terrestrial users. To compute the distance between satellite APs and terrestrial users, we adopt the Earth-centered, Earth-fixed (ECEF) coordinate system as described in \cite{mao2024intelligent}. The transformation from latitude, longitude, and altitude (LLA) coordinates to ECEF coordinates follows the model in \cite{bowring1976transformation}, i.e.:
\begin{align}
   \lv= \begin{bmatrix}
        x\\y\\z
    \end{bmatrix}=
    \begin{bmatrix}
        (N+H)\cos{B}\cos{L}\\(N+H)\cos{B}\sin{L}\\
        [N(1-e^2)+H]\sin{B}
    \end{bmatrix},\label{LLAtoECEF}
\end{align}
where $(B, L, H)$ represents the latitude, longitude, and height, respectively; $e=\frac{\sqrt{\alpha^2-\beta^2}}{\alpha}$ denotes the first eccentricity, while $\alpha$, $\beta$ are the semi-major and semi-minor axes of the Earth reference ellipsoid; $N$ denotes the prime vertical radius of curvature, i.e.: $N=\frac{\alpha}{\sqrt{1-e^2\sin^2{B}}}$. Since the connectivity between the satellite APs and terrestrial users depends not only on distance but also on the elevation angle, a user can be served by a satellite only if the elevation angle exceeds a predefined threshold. Otherwise, the connection is considered infeasible.
For evaluating the elevation between the satellite AP and the terrestrial user, we convert the ECEF coordinates $(x,y,z)$ of the satellites into east-north-up (ENU) coordinates \cite{hofmann2012global} based on the LLA position of the terrestrial user, i.e., 
\begin{equation}
\begin{split}
    \text{EAST} &= -\Delta x\sin{L}+\Delta y\cos{L},\\
    \text{NORTH} &= -\Delta x\sin{B}\cos{L} \\
    &-\Delta y\sin{B}\sin{L}+\Delta z\cos{B}, \\
    \text{UP} &= \Delta x\cos{B}\cos{L} \\
    &+\Delta y\cos{B}\sin{L}+\Delta z\sin{B},\\
\end{split} 
\end{equation}
where $\Delta x, \Delta y,$ and $\Delta z$ represent the positional differences in ECEF coordinates between the satellite and the terrestrial user.
\vspace{-0.5cm}
\subsection{Communication Model}
To achieve the objectives of ISCPT in multi-user SAGINs, the transmit signal from the AP $m$ is: $s_m = \sum_{k\in\Kc_C}s_{mk}x_k+\one_{m=m_t}s_{m_ts}x_s$, 
{where $\sum_{k\in\Kc_C}|s_{mk}|^2+|s_{ms}|^2=1$ is the precoder for communication and sensing functionalities, and we define $\Kc_C=\Kc_{C, S}\cup \Kc_{C, T}$ as the communication user set, including both terrestrial and the satellite users. In this paper, a single AP is selected for the sensing target \cite{he-gc}, which produces $|s_{ms}|^2\neq 0$ when $m=m_t$; otherwise, $|s_{ms}|^2=0$, where $m_t$ denotes the selected AP for the sensing target. The signal received by the $k$-th communication user, including satellite or terrestrial user, can be represented by:}
\begin{align}
    y_k=&\underbrace{\sum_{m=1}^{M} \sqrt{\rho_m}h_{mk}s_{mk}x_{k}}_{\text{Desired signal}} +\underbrace{\sqrt{\rho_{m_t}}h_{m_tk}s_{m_ts}x_s}_{\text{Interference from the sensing signal}}\notag\\&+\underbrace{\sum_{m=1}^{M}\sum_{k^{'}\neq k, k\in\Kc_C}\sqrt{\rho_m}h_{mk}s_{mk^{'}}x_{mk^{'}}}_{\text{Interference from the communication users}}+\underbrace{n_k}_{\text{Noise}},
\end{align}
where $h_{mk}$ is the channel between the $m$-th AP and the $k$-th user, $\rho_m$ is the transmit power for each AP, $n\sim\CN(0,\sigma^2)$ denotes the additive white Gaussian noise (AWGN). The channel $h_{mk}$ is written as: 
\begin{align}
    h_{mk}=&w_{mk}\sqrt{G_BG_U}e^{-\j\frac{2\pi}{\lambda}d_{mk}},
\end{align}
{where $w_{mk}^2$ denotes the path loss, $\lambda =\frac{c}{f}$ is the wavelength, while $c$ is the speed of light; $f$ denotes the carrier frequency, while $G_B$ and $G_U$ denote the antenna gains from the AP and user sides, respectively.} Due to differences in size and hardware characteristics between terrestrial and satellite equipment, distinct antenna gains are employed. The specific parameters are provided in Section \ref{sec:simulation}.
In the context of the non-terrestrial system, we define the path loss by focusing solely on the line-of-sight (LoS) channel, i.e.:
\begin{align}
    w_{mk}^2 =    
    &20\log\left(\frac{4\pi d_{mk}}{\lambda}\right), 
\end{align}
where $k\in\Kc\triangleq\Kc_{C}\cup\Kc_P\cup\Kc_{\text{Sen}}$, and there are in total $K=|\Kc|$ users for communication, sensing, and charging; $d_{mk}$ is the distance between the $m$-th AP satellite and the $k$-th user. By leveraging the coordinate transformation from the LLA to ECEF based on \eqref{LLAtoECEF}, $d_{mk}$ is then: 
\begin{align}
    d_{mk} = \norm{\lv_m-\lv_k}
\end{align}
Therefore, the sum rate is adopted as one of the objectives in the ISCPT system to evaluate communication performance, which is given by:
\begin{align}
    R_c = \sum_{k\in\Kc_C}\log(1+\gamma_k),
\end{align}
where $\gamma_k$ is the SINR of the $k$-th user, given by 
\begin{align}
    &\gamma_k = 
    \frac{|\sum_{m=1}^{M}\sqrt{\rho_m}h_{mk}s_{mk}|^2}{\sum_{k^{'}\in\Kc_C,k^{'}\neq k}|\text{IC}_{k^{'}}|^2+|\sqrt{\rho_{m_t}}h_{m_tk}s_{m_ts}|^2+\sigma^2},
\label{com-metrics}
\end{align}
where the interference from communication user $k^{'}$ is:
\begin{align}
&\text{IC}_{k^{'}}=\sum_{m=1}^{M}\sqrt{\rho_m}h_{mk}s_{mk^{'}}.  
\end{align}

\vspace{-0.5cm}

\subsection{Sensing Model and Wireless Power Transfer}

The satellite yielding the highest sensing SINR is selected for PNT. Note that PNT using a single satellite AP is commonly considered, since multiple time-of-arrival (ToA) measurements are feasible because the satellite moves dynamically along its orbit, providing varying observation points for ToA collection information \cite{zheng2025near, shen2020singleloc}. Since the measurement error depends on the sensing SINR, we analyze the sensing SINR as follows. 

During the sensing procedure, each AP employs a distinct precoder, which leads to interference and degrades the localization performance. Consequently, the signal received at the sensing target is given by:
\begin{align}
    y_s = \underbrace{\sqrt{\rho_{m_t}}h_{m_ts}s_{m_ts}x_s}_{\text{Desired sensing signal}}+\underbrace{\sum_{k\in\Kc_C}\sum_{m=1}^{M}\sqrt{\rho_m}h_{ms}s_{mk}x_k}_{\text{Interference from communication users}}+\underbrace{n}_{\text{Noise}},
\end{align}
where $m_t$ is the selected AP for sensing. Regarding the sensing objective, we adopt the sensing SINR as the evaluation performance metric, which is \cite{he-rss, he2025RSS}:
\begin{align}
    \gamma_s = \frac{\rho_{m_t}|h_{m_ts}s_{m_ts}|^2}{\sum_{k\in\Kc_C}|\sum_{m=1}^M\sqrt{\rho_{m}}h_{ms}s_{mk}|^2+\sigma^2}. 
\label{sense-metrics}
\end{align}
In this work, we aim to maximize the sensing SINR, which serves as a key indicator of localization performance \cite{liu2022survey}. The details on satellite-based target localization can be found in \cite{so-loc}. 
Based on the communication and sensing models in \eqref{com-metrics} and \eqref{sense-metrics}, the power received by the charging users is: 
\begin{align}
    &E_p = \notag\\
    &\sum_{k_p\in\Kc_P}\bigg(\underbrace{\sum_{k\in\Kc_C}|\sum_{m=1}^M\rho_{m}h_{mk_p}s_{mk}|^2}_{\text{Power from communication users}}+\underbrace{|\rho_{m_t}h_{m_tk_p}s_{m_ts}|^2}_{\text{Power from the sensing user}}\bigg).\label{power-matrices} 
\end{align}

\section{Topological Structure Model}
\label{sec:topo}
In the multi-user SAGIN system, satellite APs connect to multiple satellite and terrestrial users via wireless links. However, due to obstructions and long propagation distances, each user can only connect to a subset of available APs. This sparse connectivity structure makes it feasible to analyze the system by leveraging its underlying topology.

\vspace{-0.4cm}

\subsection{Bipartite Graph Construction}
For the sake of clarity, we define the bipartite graph as $\Gc=(\Mc,\Kc,\Ec)$, where $\Mc$ denotes the set of APs, $\Kc$ denotes the set of all users, while $\Ec=e(m,k)$ denotes the edges. To determine the sparse connectivity of the joint network, we evaluate link availability from two key aspects: 1) visibility and 2) channel strength \cite{radhakrishnan2016survey}. When a connection is unavailable due to the lack of visibility, the corresponding channel is treated as nonexistent in both the topological graph and the actual system, effectively represented as a ``zero link''. In contrast, connections that are visible but exhibit weak channel strength are excluded from the topological graph. However, they may still exist in the physical system and potentially cause interference to other users. The following section provides a detailed definition of how connectivity is determined.

\begin{figure}
    \centering
    \includegraphics[width=0.95\linewidth]{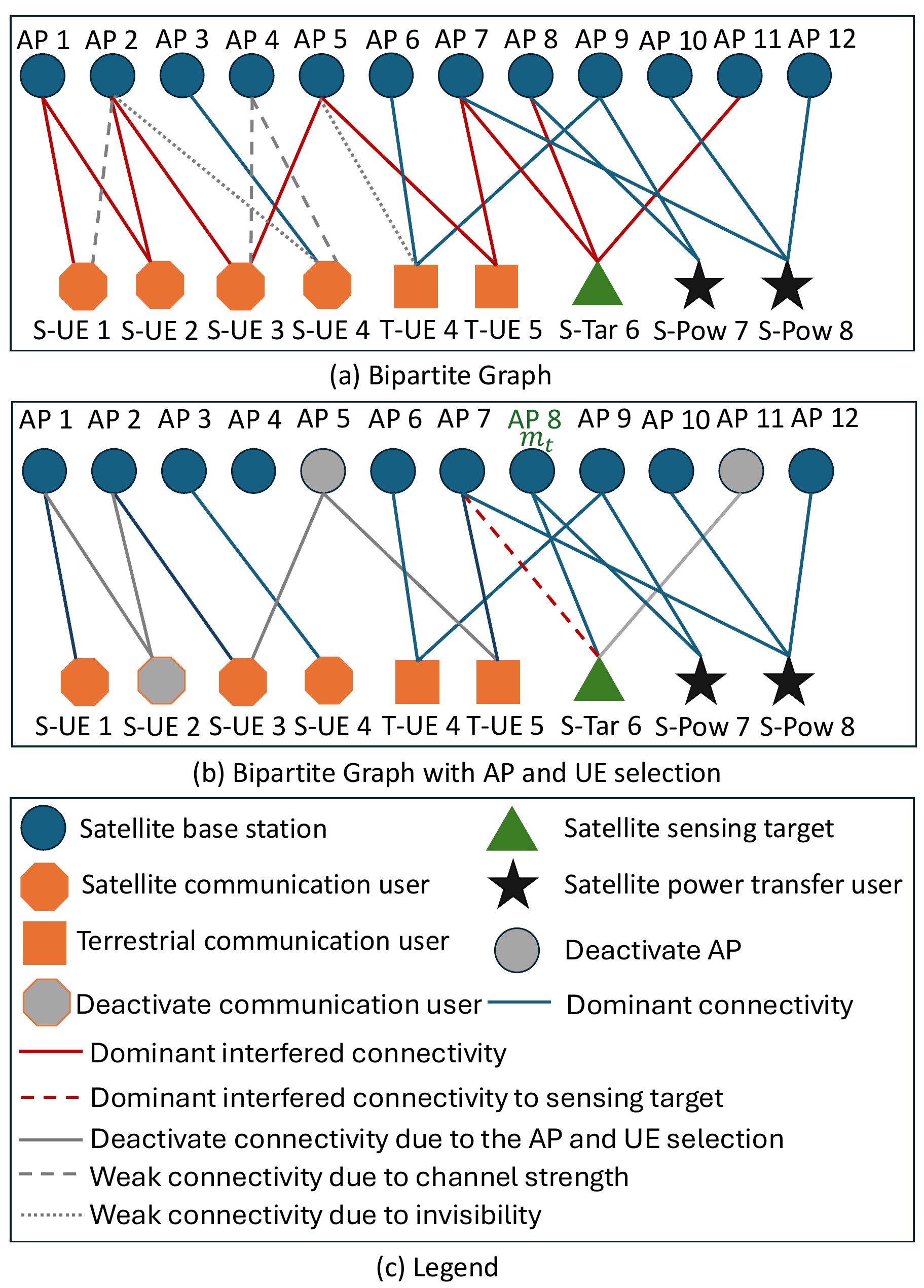}
    \caption{{(a) illustrates bipartite graph with connection between APs and users. (b) illustrates result involving the AP and user selection. Blue lines indicate no interference links, while red solid lines denote links that are impacted by interference from other nodes. Sensing target may experience interference from communication users, as shown by red dashed line. Gray solid lines depict disappeared dominant links that result from deactivating nodes, while gray nodes indicate that nodes are deactivated. Grey dashed lines denote weak connectivity due to the weak channel strength, while grey dotted line present weak connectivity due to invisibility.}}
    \label{fig:BipartiteGraph}
    \vspace{-10pt}
\end{figure}

\begin{itemize}
\item \textit{Visibility of Satellite-Terrestrial Users:} The visibility between a satellite and a terrestrial user is determined by the elevation angle, $\phi_{mk}$, which can be computed through the transformation of ENU coordinates, i.e.:
\begin{align}
    \phi_{mk} = \arctan\left({\frac{\text{UP}}{\sqrt{\text{NORTH}^2+\text{EAST}^2}}}\right). 
\end{align}
For the terrestrial user, we have
\begin{align}
    \!\!\!\! 
    e(m,k) = \begin{cases}
        &1, \quad \phi_{mk}\geq \tau_{\phi}, \ m\in\Mc, k\in\Kc_{C,T}\\
        &0,  \quad \phi_{mk}< \tau_{\phi}, \ m\in\Mc, k\in\Kc_{C,T}
    \end{cases}
\end{align}
where $\tau_\phi$ is the elevation angle threshold. 
\item \textit{Visibility of Satellite-Satellite Users:} The visibility condition of the ISL is determined by the minimum distance from the line connecting two satellites to Earth, which is:
\begin{align}
    d_{mk,e}^{\min} = \frac{\sqrt{\Norm{\lv_m}\Norm{\lv_k}-(\lv_m^\T\lv_k)^2}}{\norm{\lv_m-\lv_k}}.
\end{align}
Intuitively, if the shortest distance from the line connecting two satellites to the Earth's center is less than the Earth's radius, the direct LOS between them is obstructed by the Earth; otherwise, the satellites are mutually visible. Based on this criterion, the ISL visibility condition is:
\begin{align}
    \!\!\! e(m,k) = 
    \begin{cases}
    &1, \quad d_{mk,e}^{\min} \geq R_{\text{Earth}}, \ m\in\Mc, k\in\Kc_{S}\\
    &0,  \quad d_{mk,e}^{\min} < R_{\text{Earth}}, \ m\in\Mc, k\in\Kc_{S}
    \end{cases}
\end{align}
where $R_{\text{Earth}}=6378$ km denotes Earth radius, and we define the satellite user set as $\Kc_S\triangleq\Kc_{C,S}\cup\Kc_{P}\cup\Kc_{\text{Sen}}$, and $K_S = |\Kc_S|$.
\item \textit{Channel Strength:} This condition is used to determine the weak link by comparing the path loss with the defined threshold. For the ISL edge, we have
\begin{align}
   \!\!\! e(m,k) = \begin{cases}
        &1, \quad w_{mk}^2 \geq \tau_{\text{PL}}, \ m\in\Mc, k\in\Kc_{S}\\
        &0,  \quad w_{mk}^2 < \tau_{\text{PL}}, \ m\in\Mc, k\in\Kc_{S}
    \end{cases}
\end{align}
where $\tau_{\text{PL}}$ is the channel strength threshold. 
\end{itemize}

Consequently, a bipartite graph can be constructed, as illustrated in Fig. \ref{fig:BipartiteGraph}(a). It is seen that different shapes are used to represent APs and users. Blue edges indicate dominant links without interference, while red edges represent interference that must be considered. All colorful edges are determined by the visibility and channel strength criteria, while gray edges are ignored in the graph. Each user is connected to only a limited subset of APs, and some APs have no connectivity to any users due to the visibility constraints or weak channel conditions. In addition, it is observed that conflict links between charging users and other users do not constitute interference links. This is because no specific precoding is applied to the charging users over these links. A comprehensive explanation of a specific example that corresponds to the figure will be provided at the beginning of the next subsection. 

\vspace{-0.5cm}

\subsection{Topological Structure based Objectives Reformulation}

{Fig. \ref{fig:BipartiteGraph}(a) indicates notable interference within the system, which degrades the overall system performance, particularly in communication and sensing functionalities. It is evident that selectively deactivating certain APs or users can significantly reduce interference. For example, in Fig. \ref{fig:BipartiteGraph}(a), communication satellite users S-UEs 1, 2, 3, and T-UE 5 experience link conflicts at the AP side, leading to performance deterioration. Furthermore, the sensing target, S-Tar 6, experiences interference from communication users due to overlapping connections on APs 7, 8, and 11. It is noted that the examples of weak links are illustrated in Fig. \ref{fig:BipartiteGraph}(a), while these links are not included in the TA-framework. Consequently, the gray dashed and dotted lines are absent in Fig. \ref{fig:BipartiteGraph}(b). In addition, weak connectivity caused by invisibility is not considered in the numerical examples. However, weak connectivity due to insufficient channel strength is taken into account in the simulation study. As detailed in Section \ref{sec:SecII}, each sensing target is assigned to a single AP for the sensing task, while signals from all other APs, regardless of whether they transmit the sensing precoder, are treated as interference. The three dominant connections of S-Tar 6, $e(m_7,k_6), e(m_8,k_6), e(m_{11},k_6)$, are illustrated as red solid lines to emphasize potential interference. In addition, interference can be mitigated by deactivating selected users and APs. As shown in Fig. \ref{fig:BipartiteGraph}(b), disabling S-UE 2, AP 5, and AP 11 significantly reduces interference. This selective deactivation approach eliminates interference among communication users. For the sensing target, S-Tar 6, AP 8 has been selected as the sensing AP, i.e., $m_t=8$. However, S-Tar 6 persists in encountering interference from the communication precoder sent by AP 7 to T-UE 5, as indicated by the red dashed line in this figure. This interference with the sensing target remains unresolved. Nevertheless, the link $e(m_7,k_6)$ does not interfere with the associated communication user, T-UE 5, and our method therefore keeps AP 8 and T-UE 5 active. This can be explained by two main factors. First, since AP 8 is not selected as the sensing AP, it does not transmit the sensing precoder or sensing signal to S-Tar 6. Hence, no sensing interference is introduced over the link $e(m_7,k_5)$. Furthermore, keeping this AP-user pair active does not significantly degrade the sensing SINR and still ensures that it remains above the prescribed threshold. Thus, based on the above discussion, the precoder can be designed by controlling: 1) the user activation status, 2) the AP activation status, and 3) the AP selection for sensing.}

The objectives can be re-expressed based on the topological structure and the corresponding bipartite graph. 
Before presenting the formulation, we first introduce fundamental concepts related to bipartite graphs. 
The adjacency matrix of the bipartite graph $\Gc$ is defined as $\Am$. If a connection exists, i.e., $e(m,k)=1$, then we have $[\Am]_{m,k}=1$. {In matrix $\Am$, the first $K_C$ columns correspond to the communication connections; the $K_C+1$-th column denotes the sensing target connections, since we only consider single sensing target; and the last $K_P$ columns represent the connections of charging users.  Since we focus on the dominant connections as stated in Section \ref{sec:SecII}, we define the weight matrix $\Wm$ to denote the dominant path loss, and each element of $\Wm$ is defined as $\hat{w}_{mk}=w_{mk}[\Am]_{m,k}$. 

It is noted that this paper focuses on single sensing target scenarios, and our objective is to investigate the interference interactions among multiple functional users. Considering multiple sensing targets would increase the algorithmic complexity, while providing limited additional insight into the proposed method. Therefore, the single-target assumption is adopted to simplify the analysis while still capturing the key interference relationships among the considered functionalities. In addition, the proposed TA-based framework is inherently flexible and can be extended to more complex application scenarios. In the graph representation, both sensing targets and communication users can be naturally modeled as nodes with configurable performance requirements. Hence, the developed framework can be readily extended to scenarios involving multiple sensing targets, since such targets can be incorporated into the topology graph as nodes with their corresponding requirements, without fundamentally changing the overall framework. In future work, we will further investigate multi-target sensing scenarios, together with more sophisticated functional requirements and system configurations.}

In this paper, maximum ratio transmission (MRT) is employed in order to provide analytical clarity and actionable insights for algorithm and strategy design. It is noted that the proposed algorithm and framework can be extended to other precoding schemes, such as zero-forcing and minimum mean squared error, allowing performance to be assessed from various perspectives. By ignoring transmit power normalization, the objectives can be approximated under the topological structure based on AP and user selection, i.e.:
\begin{align}
&R_{c,\text{T}} = \sum_{k\in\Kc_C}  u_k R_{k,\text{T}}=\sum_{k\in\Kc_C}u_k\log(1+\gamma_{k,\text{T}})\label{rate-topo}\\
&\gamma_{k,\text{T}}=\notag\\
&\frac{\sum_{m=1}^{M}v_m\hat{w}_{mk}^2}{\sum_{m=1}^{M}(\sum_{k^{'}\neq k}u_{k^{'}}v_m{\hat{w}_{mk}\hat{w}_{mk^{'}}}+v_m^s{\hat{w}_{mk}\hat{w}_{ms}})+\sigma^2},\label{SINR-Topo}\\
&\gamma_{s,\text{T}}=\frac{\sum_{m=1}^{M}v_m^s\hat{w}_{ms}^2}{\sum_{k\in\Kc_C}u_{k}\sum_{m=1}^Mv_m{\hat{w}_{mk}\hat{w}_{ms}}+\sigma^2},\label{Sensing-Topo}\\
&E_{p,\text{T}}=\notag\\
&\sum_{k_p\in\Kc_P}\sum_{m=1}^M\left(\sum_{k\in\Kc_C}u_{k}v_m{\hat{w}_{mk_p}\hat{w}_{mk}}+v_m^s{\hat{w}_{mk_p}\hat{w}_{ms}}\right),\label{Power-Topo}
\end{align}
where $R_{c,\text{T}},\gamma_{k,\text{T}}$, $\gamma_{s,\text{T}}$, and $E_{p,\text{T}}$ are the topology-based sum rate, communication and sensing SINR, as well as the total energy received for the charging users. Binary parameters $u_{k}, v_m, v_m^s\in\{0,1\}$ are utilized to control the user, AP status, and the selection of the sensing AP. Regarding $u_{k}$ and $v_m$, the value of `0' indicates a deactivated status, while `1' denotes an activated status. $v_m^s$ represents the selection of the $m$-th AP for sensing. Given that only one AP is selected for the sensing target, the constraints are established as $\sum_{m=1}^M v_m^s=1, v_m^s\leq v_m$. 
{It is noted that (18)-(21) are not intended to derive equivalent formulations of (8), (11), and (12), yet providing topological representations of the corresponding objectives. The purpose is to capture the structural effects induced by the topology, such as the desired signal contribution and the resulting interference relationships, thereby facilitating strategy design and decision-making. Accordingly, power allocation and normalization are omitted in (18)-(21) for simplicity. In Section \ref{sec:simulation}, the MRT precoder with proportional power allocation is adopted to evaluate the system performance.}

\vspace{-0.2cm}

\section{Optimization Problem}
\label{sec:OptPro}
As previously stated, the multi-user SAGIN aims to meet the requirements of communication, sensing, and power transfer, which is a multi-objective optimization problem as  
\begin{align}
    \max \ &R_{c} \qquad \max\ \gamma_{s} \qquad \max\ E_{p}\\
    \st \ &\rho_m\leq \rho_{\max}.
\end{align}
Thanks to the topology-based objective re-presentation showing the optimization under the topological structure as 
\begin{align}
    \max \ R_{c,\text{T}} \qquad \max\ \gamma_{s,\text{T}} \qquad \max\ E_{p,\text{T}}. 
\end{align}
Notably, the power constraint can be eliminated in this topology-based optimization, as the power allocation is embedded within the TA precoder design.  The satellite power constraints are implicitly imposed with overall power limits, and the actual power is evently allocated according to the specified topology. In the simulation section, the power can be allocated equally, referred to as average power allocation, or proportionally, known as proportional power allocation. 
 
This optimization involves three objectives: the first objective refers to the communication metric, specifically the sum rate of all activated communication users; the second relates to the quality of the sensing target; and the third evaluates the received energy for power transfer users. The communication and sensing purposes have been defined as the main tasks; thus, we represent the multi-objective optimization into two sub-optimizations, considering the power transfer as the constraint. Therefore, the tradeoff between the sensing and communication with considering the power transfer can be balanced by the alternating optimization \cite{deb2001multi}: 
\begin{subequations}
\begin{align}
{(\Pc_1)} \qquad \max&\sum_{{k}\in\Kc_C}u_{k}R_{{k},\text{T}}\\
\st\ &\gamma_{s,\text{T}}\geq\tau_{s}, \label{Sensing-Con} \\
&E_{p,\text{T}}\geq \tau_p \label{Power-Con}, 
\end{align}    
\end{subequations}
\begin{subequations}
\begin{align}
{(\Pc_2)} \qquad \max & \ \gamma_{s,\text{T}}\\
\st\ &\sum_{{k}\in\Kc_C}u_{k}R_{{k},\text{T}}\geq\tau_c, \\
&E_{p,\text{T}}\geq \tau_p, 
\end{align}    
\end{subequations}
where $\tau_s, \tau_p, \tau_c$ are the performance thresholds for sensing, power, and communication, respectively. One challenge is to determine the thresholds for three objectives. In the subsequent procedure, we further reformulate these thresholds to make them more tractable and easier to control.

Regarding the communication performance, as proven in \cite{khalilsarai2018fdd}, the multiplexing gain of the sum rate can be converted to the maximum cardinality matching problem within the bipartite graph, taking into account the AP and the user-selected subgraph. The subgraph is defined as: $\Gc^{'}=(\Mc^{'},\Kc_C^{'},\Ec^{'})$, where $\Mc^{'}$ and $\Kc_C^{'}$ represent the selected AP and communication node sets, respectively, and $\Ec^{'}$ comprises the edges associated with these selected nodes. Furthermore, as discussed in \cite{yu2021active}, the communication SINR for each selected user shall also be considered as a constraint. 
Thus, the sub-optimizations are approximately by: 
\begin{subequations}
\begin{align}
{(\Pc_1^{'})} \qquad \max\ &|\Mc(\Mc^{'},\Kc_C^{'})|\label{Objective-P1}\\
\st\ &\gamma_{k,\text{T}}\geq\tau_c, \quad \forall \ k\in\Kc_C^{'}\label{SINR-P1}\\
&\gamma_{s,\text{T}}\geq\tau_s,\label{Sensing-P1}\\
&E_{p,\text{T}}\geq \tau_p,
\label{Power-P1}
\end{align}    
\end{subequations}
\begin{subequations}
\begin{align}
{(\Pc_2^{'})} \qquad \max\quad& \gamma_{s,\text{T}}\\
\st\ &|\Mc(\Mc^{'},\Kc_C^{'})|\geq N_{\tau},\\
&\gamma_{k,\text{T}}\geq \tau_c, \quad \forall\ k\in\Kc_C^{'},\\
&E_{p,\text{T}}\geq \tau_p, 
\label{P2_1}
\end{align}    
\end{subequations}
where $|\Mc(\Mc^{'},\Kc_C^{'})|$ denotes the maximum cardinality bipartite matching number of the selected subgraph. This subgraph includes communication user nodes alone, as only these users have the binary selection.
{The following illustrates the derivation of sub-optimizations to the MILP problem, where the thresholds $\tau_c, \tau_p, \tau_s$ are not used for the optimization process.}

\vspace{-0.2cm}

\subsection{TA-Sum Rate Optimization ($\Pc_1$)}

The sub-optimization problem $(\Pc_1^{'})$ can be approximated as a MILP $(\Pc_1^{''})$, and the solution is feasible for the optimization problem $(\Pc_1^{'})$. The MILP can be obtained as:   
\begin{subequations}
    \begin{align}
(\Pc_1^{''})&\max_{u_k,v_m,v_m^s,z_{m,k}}\quad \sum_{m=1}^M\sum_{k\in\Kc_C}z_{m,k}\label{Objective-v2}\\
   \st\quad& u_k\leq\sum_{m=1}^M [\Am]_{m,k}v_m, \quad\forall\ k\in\Kc_C,\label{P1-UE-v2}\\
   &v_m\leq\sum_{k\in\Kc_C}[\Am]_{m,k}u_k+[\Am]_{m,K_C+1}\notag\\&\qquad+\sum_{k_p=1}^{K_P}[\Am]_{m,K_C+1+k_p},\qquad\quad\forall\ m\in\Mc\label{P1-AP-v2}\\
   &z_{m,k}\leq[\Am]_{m,k}, \quad\forall\ k\in\Kc_{C}, m\in \Mc\label{Matching-1-v2}\\
   & \sum_{k\in\Kc_C}z_{m,k}\leq v_m,\quad\forall\ m\in\Mc\label{Matching-2-v2}\\
   & \sum_{m=1}^Mz_{m,k}\leq u_k,\quad\forall\ k\in\Kc_C\label{Matching-3-v2}\\
   &\hat{w}_{mk}^2+c_1(1-u_k)+c_2(1-v_m)\geq\notag\\
   &\left(\sum_{k^{'}\in\Kc_C}u_{k^{'}}\hat{w}_{mk^{'}}\hat{w}_{mk}+v_m^s\hat{w}_{ms}\hat{w}_{mk}\right)\hat{\tau}_{c,m},\notag\\
   & \qquad\qquad\qquad\qquad\qquad\quad\forall \ k\in\Kc_C,\ m\in\Mc\label{SINR-v2}\\
&\sum_{k_p\in\Kc_P}\sum_{m=1}^Mv_m\hat{w}_{mk_p}\geq \hat{\tau}_p\sum_{k_p\in\Kc_P}\sum_{m=1}^M\hat{w}_{mk_p}, \label{Power-v3} \\
&\sum_{m=1}^Mv_m^s\hat{w}_{ms}^2+c_4(1-u_k)\geq\notag\\
&\qquad\hat{\tau}_{s,k}\left(\sum_{m=1}^Mv_m^s\hat{w}_{ms}^2+\sum_{m=1}^Mv_m\hat{w}_{mk}\hat{w}_{ms}\right), \notag\\
&\qquad\qquad\qquad\qquad\quad\qquad\qquad\qquad    \forall \ k\in\Kc_C\label{Sensing-1-v2}\\
&v_m^s\leq v_m, \quad \forall \ m\in\Mc\label{Sensing-2-v2}\\
  &\sum_{m=1}^M [\Am]_{m,K_C+1}v_m^s=1, \label{Sensing-3-v2}\\
  &u_k, v_m, v_m^s\in\{0,1\}, \ \forall k\in\Kc_C, m\in\Mc \label{Binary}\\
  &z_{m,k}\in[0,1],\ \forall k\in\Kc_C, m\in\Mc \label{constriantZ}
    \end{align}
    \label{P1}
\vspace{-0.5cm}
\end{subequations}

\noindent where $\hat{\tau}_{c,m},\hat{\tau}_{p},\hat{\tau}_{s,k}\in[0,1]$ are thresholds for controlling the communication SINR, power transfer, sensing SINR, respectively. 
Here, $c_1,c_2,c_3,$ and $c_4$ are sufficiently large constants, and $z_{m,k}$ can be relaxed from binary value, i.e., $z_{m,k}=\{0,1\}$ to the continuous one $[0,1]$ \cite{khalilsarai2018fdd}.\\
Proof: See Appendix \ref{sec:AppA}. 

\begin{remark}
For communication, the sum-rate can be decomposed into two parts: the maximum matching problem, which determines the prelog factor, and the SINR constraint, ensuring the required communication quality of service (QoS). Specifically, the objective function in \eqref{Objective-v2} characterizes the maximum matching cardinality. Constraint \eqref{P1-UE-v2} specifies that each activated communication user shall be associated with at least one AP, while constraint \eqref{P1-AP-v2} requires that each activated AP serves either one activated communication user, one sensing target, or one charging user. The matching relationships are further governed by the constraints in \eqref{Matching-1-v2}–\eqref{Matching-3-v2}. The communication SINR requirement is captured by constraint \eqref{SINR-v2}. 
The charging limitations from the constraint \eqref{Power-Con} are enforced through constraint \eqref{Power-v3}. Finally, constraints \eqref{Sensing-1-v2}–\eqref{Sensing-3-v2} are derived from the sensing requirement \eqref{Sensing-Con}.  
\end{remark}

The topological structure-based expression provides valuable insights regarding each objective through the constraints. Constraint \eqref{SINR-v2} indicates that for each dominant connection of each user, the desired power shall exceed the total received power by a specified percentage, $\hat{\tau}_{c,m}$. Otherwise, the relevant AP or user should be deactivated. In addition, this constraint illustrates the relationship of communication interference based on specific decisions regarding the user and AP status. In this context, we utilize sufficiently large constants $c_1$ and $c_2$ to demonstrate the conditions under which the user and the AP are activated, ensuring the linearization of the SINR constraint. For instance, when $u_k=1$ and $v_m=0$, indicating that the user is activated while the AP is deactivated, the constraint shall be satisfied. Conversely, when $u_k=1$ and $v_m=1$, it is necessary to determine whether the interference meets with the specified percentage, $\hat{\tau}_{c,m}$. Different from the communication and sensing functions, the charging power primarily relies on the status of the AP. Therefore, \eqref{Power-v3} represents the received power by all charging users in relation to the corresponding AP status. The threshold $\hat{\tau}_p\in[0,1]$ indicates the minimum amount of power that can be transferred from APs. Specifically, `0' indicates that the charging function is entirely unnecessary, while `1' means that all dominant connections of the charging user are required. The sensing SINR constraint in \eqref{Sensing-1-v2} is employed to select the sensing AP by comparing the interference connections for each communication user. The constraint \eqref{Sensing-3-v2} states the requirement of selecting one existing dominant connection to the sensing target. 

\subsection{TA-Sensing SINR Optimization $(\Pc_2)$}

The sub-optimization problem $(\Pc_2^{'})$ can be approximately transferred as:
\begin{subequations}
\begin{align}
(\Pc_2^{''})& \min_{u_k,v_m,v_m^s,t} \ t\label{Objective-P2}\\
\st\quad&u_k\leq\sum_{m=1}^M [\Am]_{m,k}v_m, \quad\forall\ k\in\Kc_C,\label{P2-UE-v2}\\
   &v_m\leq\sum_{k=1}^{K_C}[\Am]_{m,k}u_k+[\Am]_{m,K_C+1}\notag\\
   &\qquad+\sum_{k_p=1}^{K_P}[\Am]_{m,K_C+k_p+1},\qquad\quad\forall\ m\in\Mc\label{P2-AP-v2}\\ &\sum_{m=1}^Mv_m\hat{w}_{mk}\hat{w}_{ms}\leq t\hat{w}_{ms}^2+c_1(1-u_k)+c_2(1-v_m^s),\notag\\
&\quad\qquad\qquad\qquad\qquad\qquad\quad\forall k\in \Kc_C,m\in\Mc \label{Sensing-1-P2}\\
&\sum_{m=1}^M [\Am]_{m,K_C+1}v_m^s=1.\label{Sensing-2-P2}\\
&v_m^s\leq v_m, \quad \forall \ m\in\Mc\label{Sensing-3-P2}\\
&\sum_{k\in\Kc_C} u_k\geq N_\tau \label{Prelog-P2}\\
&\hat{w}_{mk}^2+c_1(1-u_k)+c_2(1-v_m)\geq\notag\\
   &\qquad\left(\sum_{k^{'}\in\Kc_C}u_{k^{'}}\hat{w}_{mk^{'}}\hat{w}_{mk}+v_m^s\hat{w}_{ms}\hat{w}_{mk}\right)\hat{\tau}_{c,m},\notag\\
   & \qquad\qquad\qquad\qquad\qquad\quad\forall \ k\in\Kc_C,\ m\in\Mc\label{SINR-P2}\\
&\sum_{k_p\in\Kc_P}\sum_{m=1}^Mv_m\hat{w}_{mk_p}\geq \hat{\tau}_p\sum_{k_p\in\Kc_P}\sum_{m=1}^M\hat{w}_{mk_p}, \label{Power-P2}\\
 &u_k, v_m, v_m^s\in\{0,1\}, \ \forall k\in\Kc_C, m\in\Mc \label{Binary-P2}
\end{align}
\label{P2}
\vspace{-0.5cm}
\end{subequations}

\noindent where $N_\tau$ indicates the threshold number of activated communication users required to ensure the prelog factor. \\
Proof: See Appendix \ref{sec:AppB}. 
\begin{remark}
    The sensing objective, represented by the sensing SINR, is achieved through the objective \eqref{Objective-P2} and the constraints \eqref{Sensing-1-P2} and \eqref{Sensing-2-P2}. The prelog factor and communication QoS can be guaranteed by the constraints \eqref{Prelog-P2} and \eqref{SINR-P2}. The power transfer constraint is achieved from \eqref{Power-P2}. 
\end{remark}
{ 
In the aforementioned sub-optimizations, four thresholds can affect optimization performance, i.e., $\hat{\tau}_{c,m}, N_\tau, \hat{\tau}_{s,k}, \hat{\tau}_p$, which correspond communication, sensing, and charging objectives, respectively. Specifically, the threshold $\hat{\tau}_{c,m}$ represents the communication QoS requirement from the AP side, which indicates that the proportion of the total connection weight contributed by the activated communication users associated with AP $m$ must exceed the predefined threshold $\hat{\tau}_{c,m}$. The threshold $\hat{\tau}_{s,k}$ corresponds to the sensing SINR requirement in the topology graph, meaning that the weight between the sensing target and the selected AP must exceed a certain proportion of the total weight at the sensing target. The threshold $\hat{\tau}_p$ characterizes the WPT requirement and is defined as the proportion of the total weights between the activated APs and the selected power transfer users relative to the total weight toward all activated charging users. It is noted that $N_\tau$ represents the minimum number of served communication users required in the system. 
To communication and sensing, a higher threshold implies that links with stronger channel conditions are prioritized, which improves QoS but may reduce the number of served users. Conversely, a lower threshold allows the system to serve more users while relaxing the QoS requirement. For the WPT, the threshold $\hat{\tau}_p$ helps prevent excessive AP deactivation, thereby avoiding a reduction in the total received energy.

In the proposed algorithm, the thresholds do not need to be manually selected, which can gradually converge to near-optimal values through the alternating process. This mechanism also ensures that the proposed method can achieve a stable performance. In addition, if the system complexity does not allow multiple rounds of alternating optimization, the thresholds can still be determined according to the system requirements. We note that the thresholds $\hat{\tau}_{c,m}$, $\hat{\tau}_{s,k}$, and $\hat{\tau}_p$ lie within the bounded range $[0,1]$. These thresholds do not directly represent the absolute values of the objectives. Instead, they reflect the required performance level of each objective, such as high communication quality, instead of specific numerical objective values. 
For example, if the system prioritizes high-quality communication performance, the threshold $\hat{\tau}_{c,m}$ can be set close to or equal to $1$, which ensures that only communication links with strong channel conditions are considered.
As a result, as long as the performance requirements of the considered functions remain unchanged, the same threshold settings can generally be adopted even when the network topology varies. This property makes the devised framework relatively insensitive to the exact threshold values and enhances its robustness across different system configurations.
} The alternating algorithm used to solve the considered problem is summarized in Algorithm \ref{Algorithm1}. In this paper, we set the maximum iteration number $N_{\text{iter}}=10$. The vectors $\uv=[u_1,\dots,u_k]^\transp, \vv=[v_1,\dots,v_m]^\transp,$ and $\vv^s = [v_1^s,\dots,v_m^s]^\transp$ are the output of optimization $(\Pc_1^{''})$, while $\uv_1,\vv_1,$ and $\vv_1^s$ are that of optimization $(\Pc_2^{''})$. 

\begin{algorithm}[t]
\small
    \caption{TA Alternating  ISCPT Algorithm}
    \begin{algorithmic}[1]
        \STATE{Initialize the $[0,1]$ threshold parameters $\hat{\tau}_{c,m},\hat{\tau}_{p}$, and\\ $\hat{\tau}_{s,k}$, and the maximum iteration number $N_{\text{iter}}$.}
        \STATE{FLAG=1}
        \WHILE{FLAG=1}
        \STATE{Define the iteration parameter $i=1$}
        \STATE{Solve the optimization problem $(\Pc_1^{''})$}
        \STATE{Output of $(\Pc_1^{''})$: vectors $\uv,\vv,\vv^s$.}
        \STATE{By applying the results of optimization $(\Pc_1^{''})$, the threshold parameter of $(\Pc_2^{''})$, $\hat{\tau}_{c,m}$}, can be calculated. 
        \STATE{Solve the optimization problem $(\Pc_2^{''})$.}
        \STATE{Output of $(\Pc_2^{''})$: $\uv_1,\vv_1,\vv^s_1$.}
        \STATE{By applying the results of optimization $(\Pc_2^{''})$, the threshold parameters of $(\Pc_1^{''})$, $\hat{\tau}_{c,m},\hat{\tau}_{s,k}$, can be calculated}
        \STATE{$i=i+1$}
        \IF{thresholds remain consistent or $i\geq N_{\text{iter}}$}
        \STATE{$\text{FLAG}=0$}
        \ENDIF
        \ENDWHILE
        \STATE Output: the user, AP status, and sensing AP selection vectors: $\uv,\vv,\vv^s$.
    \end{algorithmic}
    \label{Algorithm1}   
\end{algorithm}

\subsection{Greedy Algorithm Design}
The preceding two subsections yield impressive results applying MILP and the alternating algorithm within the topological structure. However, the complexity of these algorithms, particularly the latter is significantly high, rendering them impractical for large-scale real-world systems.

{The insights obtained from the TA-based representation and the corresponding optimization formulations enable the design of the proposed GMAP-inspired greedy algorithm. Our goal is to jointly optimize the performance of the three considered objectives. From the TA-based representations in \eqref{rate-topo}-\eqref{Power-Topo}, it is observed that the system performance is determined by the relationships among the edges in the topology graph. In particular, good performance is attained when desired links are strengthened while the interference among edges is limited.
This insight is also reflected in the constraints of the optimization problems. For instance, the communication QoS constraints in \eqref{SINR-v2} and \eqref{SINR-P2} essentially control the proportion between the weight of the desired edge and the total weight of all received edges at the user. Similar structures also appear in the sensing formulations. Accordingly, the greedy algorithm is designed to select edges that favor strong desired links while limiting interference relationships in the topology graph.
Thus, these insights enable the intuitive design of a greedy algorithm through the application of the assignment problem, specifically the GMAP \cite{park1998lagrangian}. The desired link weights are treated as the `profit', while the interference relationships are modeled as the corresponding `cost'.} The final evaluation metric is then defined as the `final profit':  
\begin{align}
    &[\Pm_c]_{m,k} = 
        v_mu_{k}\hat{w}_{mk}^2, \quad m\in\Mc, k\in\Kc_C\\
    &[\pv_s]_{m} = v_m^s\hat{w}_{ms}^2, \quad m\in\Mc\\
    &[\Pm_p]_{m,k} =  v_m\hat{w}_{mk}^2, \quad m\in\Mc, k\in\Kc_P\\
    &[\Cm_c]_{m,k} = \notag\\
    &v_mu_k\left(\underbrace{\sum_{k^{'}\neq k}\hat{w}_{mk}u_{k^{'}}\hat{w}_{mk^{'}}}_{\text{Cost from communication user}}+\underbrace{v_m^s\hat{w}_{mk}\hat{w}_{ms}}_{\text{Cost from the sensing target}}\right), \notag\\
    & \qquad\qquad\qquad\qquad\qquad\qquad\quad m\in\Mc, \ k,k^{'}\in\Kc_C\label{Cost-C}\\
    &[\Cm_s]_{m,k} = v_mu_k\hat{w}_{ms}\hat{w}_{mk},\quad m\in\Mc, \ k\in\Kc_C\label{Cost-S}
\end{align}
where $\Pm=[\Pm_c,\pv_s,\Pm_p]\in\CC^{M\times K}$ refers to the profit matrix for all users. The matrix $\Cm_c\in\CC^{M\times K_C}$ indicates the communication costs for users, while $\Cm_s$ represents the sensing target costs. Subsequently, based on the profit and sensing evaluation, we can determine the final profit. Addressing these three objectives leads to the splitting of the final profit into three components as: 
\begin{align}
    [\qv^c]_k &= \frac{\sum_{m=1}^M[\Pm_c]_{m,k}}{\sum_{m=1}^M([\Pm_c]_{m,k}+[\Cm_c]_{m,k})}, \qquad\qquad k\in\Kc_C\label{FP-C}\\
    q^s &= \frac{\sum_{m=1}^M[\pv_s]_m}{\sum_{m=1}^M\sum_{k\in\Kc_C}([\Cm_c]_{m,k}+[\pv_s]_m)},\label{FP-S}\\
    q^p &= \sum_{m=1}^M\sum_{k\in\Kc_P}[\Pm_p]_{m,k},\label{FP-P}
\end{align}
where $\qv^c\in\CC^{K_C\times 1}$ represents the final profit of communication users, while the scalars $q^s$ and $q^p$ are the final profits from sensing and charging, respectively. {In the GMAP, `Profit' represents the weight of desired links, while `Cost' indicates the interference experienced by users. For communication and sensing users, the desired links are those transmitting the precoder and symbols to the users, such as the blue solid links in Fig. \ref{fig:BipartiteGraph}(b). Regarding the charging users, all links connecting to them are desirable. For the communication and sensing users, the interfered link is the one that conflicts with other communication or sensing users at the AP. The charging user does not have interference links. Thus, the dimensions of the cost matrix are: $\Cm=[\Cm_c,\Cm_s]\in\CC^{M\times (K_S+K_C)}$. The denominator of the `Final Profit' is constructed by incorporating the corresponding profit term, which ensures that the expression is well defined, i.e., the denominator is not zero. As a result, the formulated final profit maps all three components into the range $(0,1]$. By incorporating the profit term into the denominator, the threshold design can also be simplified.}
According to the mentioned `Profit', `Cost', and `Final Profit', the low complexity greedy algorithm is shown as Algorithm \ref{Algorithm2}. 
\begin{algorithm}[t]
\small
    \caption{GMAP-inspired TA Greedy Algorithm}
    \begin{algorithmic}[1]
        \STATE{Initialize the vector $\uv = \one_{K_C}, \vv=\one_{M},\vv^s=\zero_M$, adjacency matrix $\Am$,  the maximum allowable conflict connection number $N_{in}$, sensing threshold $\tau_s^g$.} 
        \IF{$\hat{\wv}_s=\zero_M$}
        \STATE{Cannot detect the target, $q_s=0,\vv^s=\zero_M$}
        \ELSE
        \STATE{Generate the sensing AP selection vector $\vv^s, v_m^s=1, m=\arg\max_m{\hat{w}_{ms}^2}$}.
        \STATE{Initialize the cost matrix and the final profit according to \eqref{Cost-S} and \eqref{FP-S} as $\Cm_s,  q^s$.}
        \WHILE{$q^s<\tau_s^g$}
        \STATE{Select the worst AP as $m_1$.}
        \STATE{Calculate the final profit if close the AP as $q^s_1$.}
        \STATE{Select the communication users, $\Kc_2$, conflict at the AP.}
        \STATE{Calculate the final profit if closing the conflict users as $q^s_2$.}
        \IF{$q_1^s>q_2^s$}
        \STATE{Update the AP status $v_{m_1}=0$.}
        \ELSE
        \STATE{Update the user status $u_k=0,k\in\Kc_2$}
        \ENDIF
        \STATE{Update the final profit $q^s$}
        \ENDWHILE
        \ENDIF
        \STATE{Check the connectivity at each AP, $N_{c,m}=\sum_{k\in\Kc_C}u_kv_m[\Am]_{m,k}$.}
        \STATE{Calculate the cost and profit matrices as $\Pm_c,\Cm_c$.}\WHILE{$\max{N_{c,m}}>N_{in}$}
        \STATE{Select the worst AP according to the cost matrix as $m_3$.}
        \STATE{Calculate the final profit if close the AP as $q^c_3$.}
        \STATE{Select the conflict users as $\Kc_4$.}
        \STATE{Calculate the final profit if close the conflict users as $q^c_4$.}
        \IF{$q_3^c>q_4^c$}
        \STATE{Update the AP status $v_{m_3}=0$.}
        \ELSE
        \STATE{Update the user status $u_k=0,k\in\Kc_4$}
        \ENDIF
        \STATE{Update the cost and profit $\Pm_c,\Cm_c$.}
        \ENDWHILE
        \STATE Output the status vectors $\uv,\vv,\vv^s$. 
    \end{algorithmic}
    \label{Algorithm2}   
\end{algorithm}

This TA-greedy algorithm can be understood through the following four points:  
\begin{itemize}
    \item {We start by evaluating the performance of the sensing target. Prior to determining the sensing performance, it is essential to verify if the target is detectable or localizable by the APs, as indicated in Line 2 of Algorithm \ref{Algorithm2}. The interference from the sensing target is solely caused by the communication users as indicated in the aforementioned formulations.} 
    \item {The core idea of this greedy algorithm is to facilitate decision-making. The AP is typically chosen based on the cost $q^s$ related to the sensing target and the number of connected users $N_{c,m}$ for communication purposes. 
    We have two options: to close the dominant AP or the conflict users regarding the dominant AP. The algorithm makes the decision for each dominant AP until the requirements are satisfied. }
    \item {The selection of the threshold impacts the performance of sensing and communication. A higher threshold $\tau_s^g$ results in reduced interference from the communication user to the sensing target. A lower value of conflict $N_{in}$ indicates greater orthogonality among users, leading to diminished interference and a lower proportion of activated users.}
    \item {The charging performance can be assessed when we approach the user and AP selection; therefore, in the aforementioned process, decisions are made based on the final profit, which serves to evaluate overall performance.}
\end{itemize}

\section{Numerical Results}
\label{sec:simulation}

This section evaluates the performance of the proposed optimization and greedy algorithms in terms of communication, sensing, and WPT. Before presenting the simulation results, we first describe the baseline simulation settings and the benchmark schemes used for comparison.

\begin{table}[t]
\centering
\caption{Simulation Parameters.}
\begin{tabular}{c|c|c|c}
\hline
\hline
Parameter & Value & Parameter & Value \\
\hline
$G_B$ & 30 dBi &$G_U$ (Satellite) & 30 dBi\\\hline
$G_U$ (Terrestrial) & 40 dBi & $f$ & 2 GHz \\\hline
$W$ & 100 MHz & $K_P$ & 4\\\hline
$N_{in}$ &1 &$\tau_s^g$ & 1\\\hline
$\rho_{\max}$ & 10 W & $\hat{\tau}_{c,m}$ & 0.3\\\hline
$\hat{\tau}_p$ & 0.5 & $\hat{\tau}_{s,k}$ & 0.3\\\hline
$\tau_\phi$ & $15^\circ$&$\tau_{\text{PL}}$ & $-190$ dB\\\hline
\end{tabular}
\label{TableI}
\end{table} 

\subsection{Simulation Setups}

This paper considers a system where LEO satellites serve as both APs and users, with the user satellites positioned in lower orbits relative to the LEO APs. Two distinct LEO constellations are constructed to represent the satellite APs and users. The AP constellation operates at an altitude of $700$ km and consists of $16$ orbital planes, while the user constellation is deployed at an altitude of $300$ km with $4$ orbital planes. Both constellations comprise $128$ satellites each, and we randomly select $K_S$ LEOs from the user constellation as the LEO users. We also consider the terrestrial users located around at $5$ remark cities: 1) Berlin $(52.52^\circ, 13.4^\circ)$; 2) New York $(40.71^\circ, -74^\circ)$; 3) London $(50.5^\circ, -0.13^\circ)$; 4) Beijing $(39.9^\circ, 116.4^\circ)$; and 5) Sydney $(-33.87^\circ, 151.21^\circ)$.
The parameters used are summarized in the Table \ref{TableI}.

We evaluate the performance of the proposed algorithms by comparing them against the following methods: TA-ISCPT alternating optimization, initialized TA-ISCPT optimization, TA-greedy algorithm, greedy user selection, and a baseline with no user/AP selection. Greedy user selection is conducted based on two key criteria:
\begin{itemize}
\item \textit{Sensing Objective:} The AP with the strongest channel to the sensing target is selected, and all non-orthogonal communication users relative to the sensing target are deactivated to mitigate interference.
\item \textit{Communication Objective:} Only orthogonal users are selected, with priority given to those with strong channel conditions.
\end{itemize}

\subsection{Simulation Results}

\begin{figure}
    \centering  \includegraphics[width=0.95\linewidth]{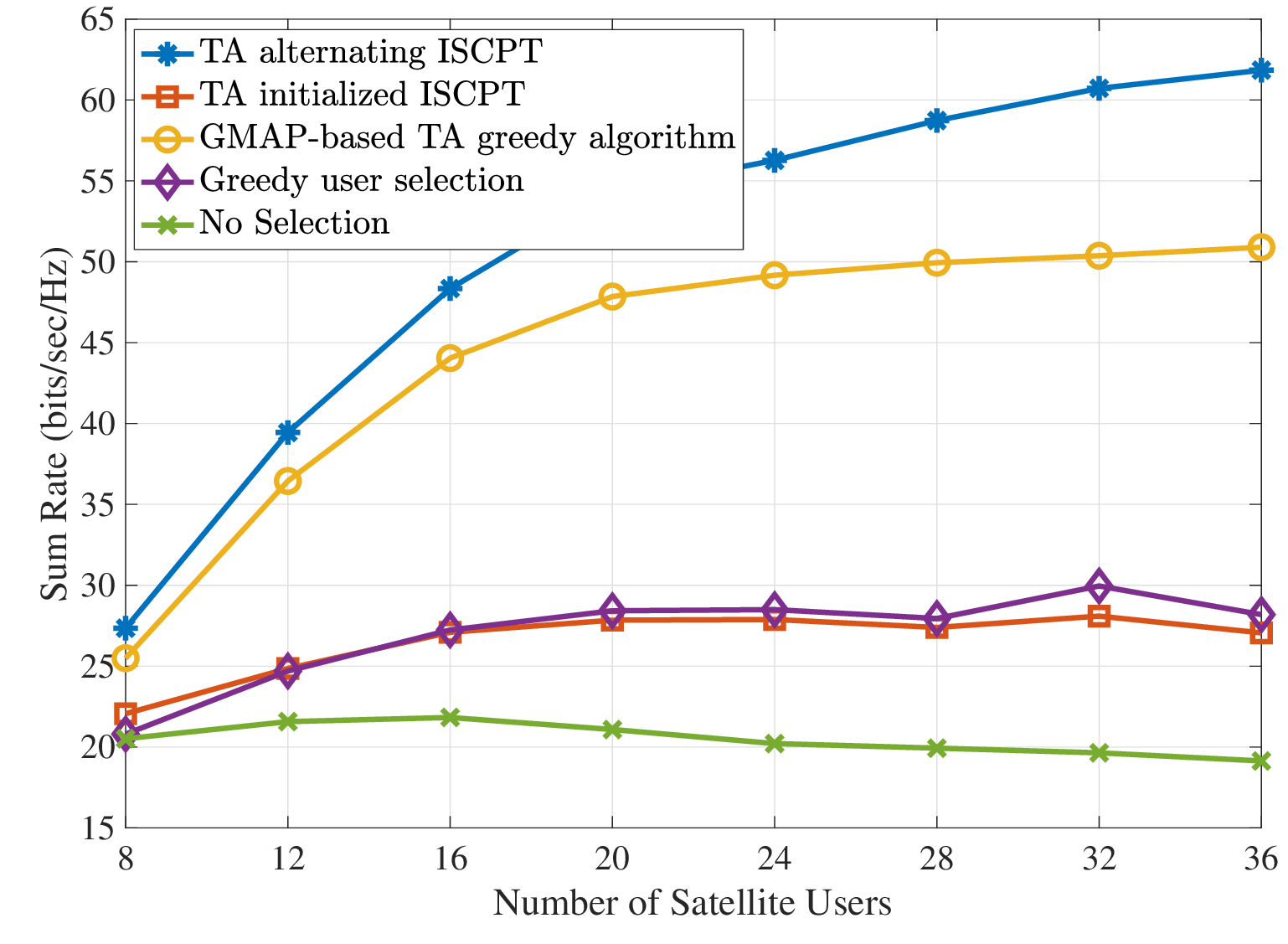}
    \caption{Sum rate versus the number of satellite users $K_S$. }
    \label{UE-1}
\end{figure}

\begin{figure}
    \centering
\includegraphics[width=0.95\linewidth]{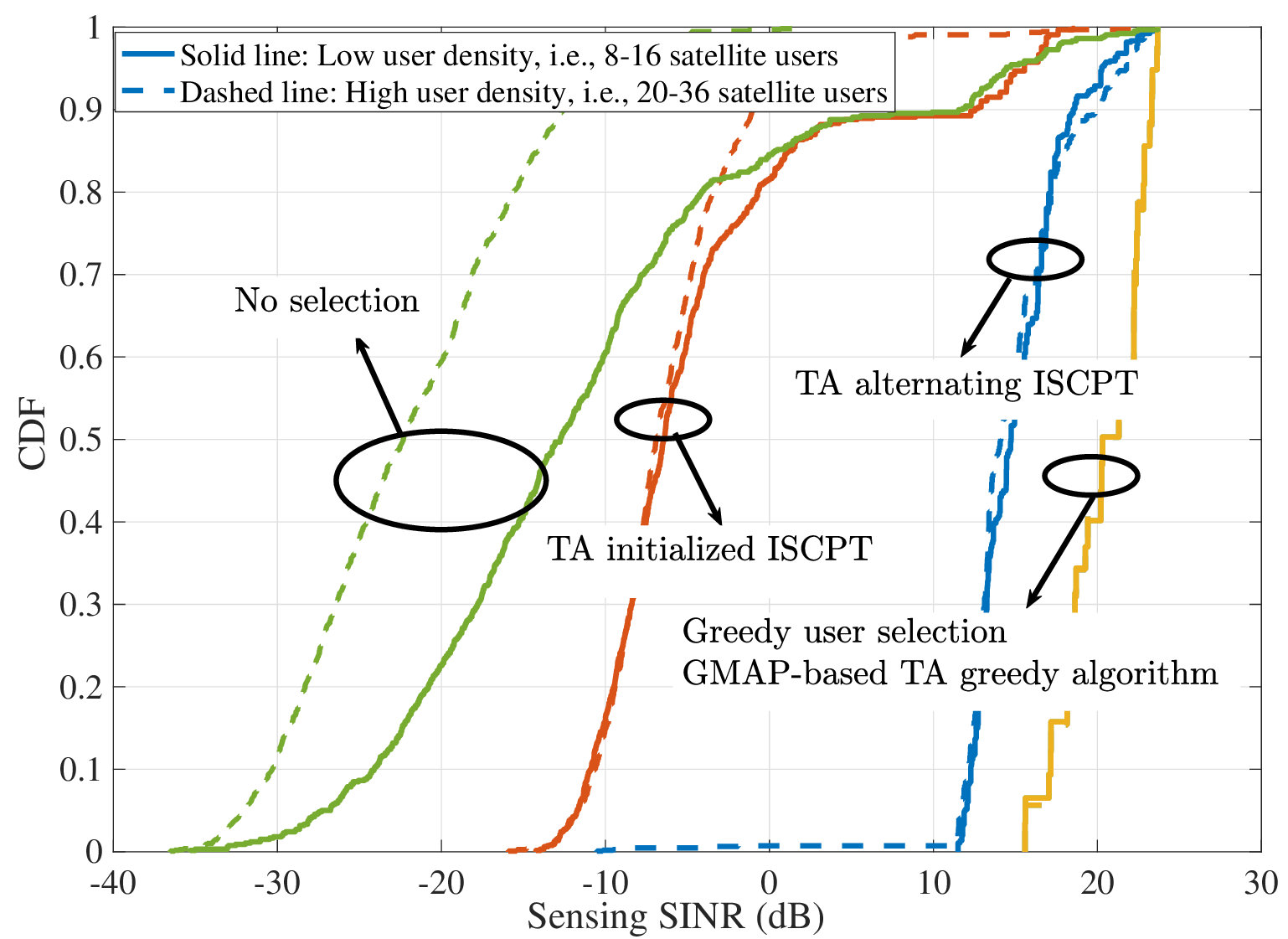}
    \caption{CDF of sensing SINR, where the solid lines is the low number of communication users ($8$-$16$ satellite users), and dashed lines denote the high number of communication users ($20$-$36$ satellite users).}
    \label{UE-2}
\vspace{-0.5cm}
\end{figure}

\begin{figure}
    \centering
\includegraphics[width=0.95\linewidth]{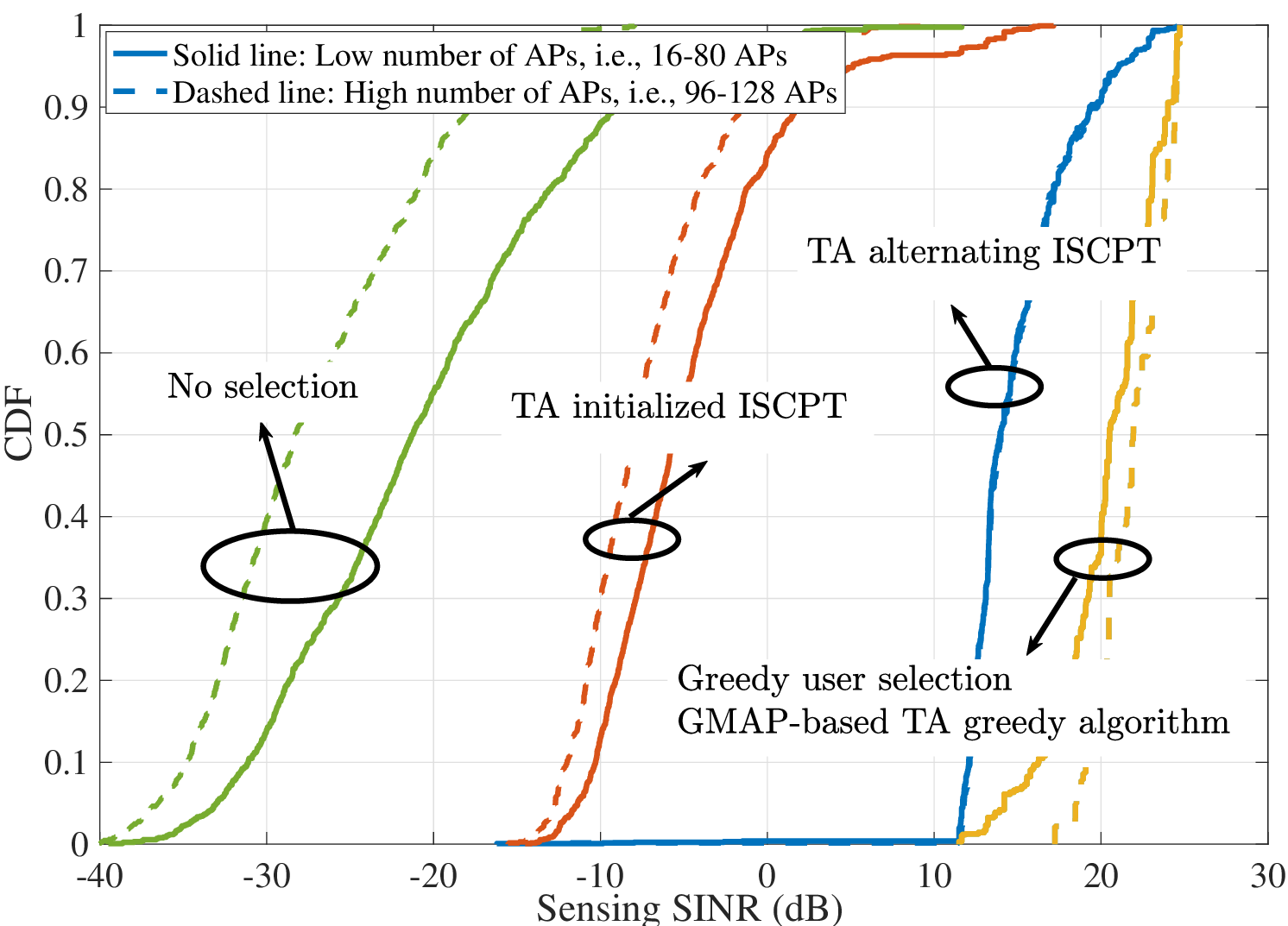}
    \caption{CDF plot of sensing SINR, where the solid lines denote the low number of satellite APs ($16$-$80$ APs), and dashed lines denote the high number of satellite APs ($96$-$128$ APs).}
    \label{AP-2}
\end{figure}

\begin{figure}
    \centering
\includegraphics[width=0.9\linewidth]{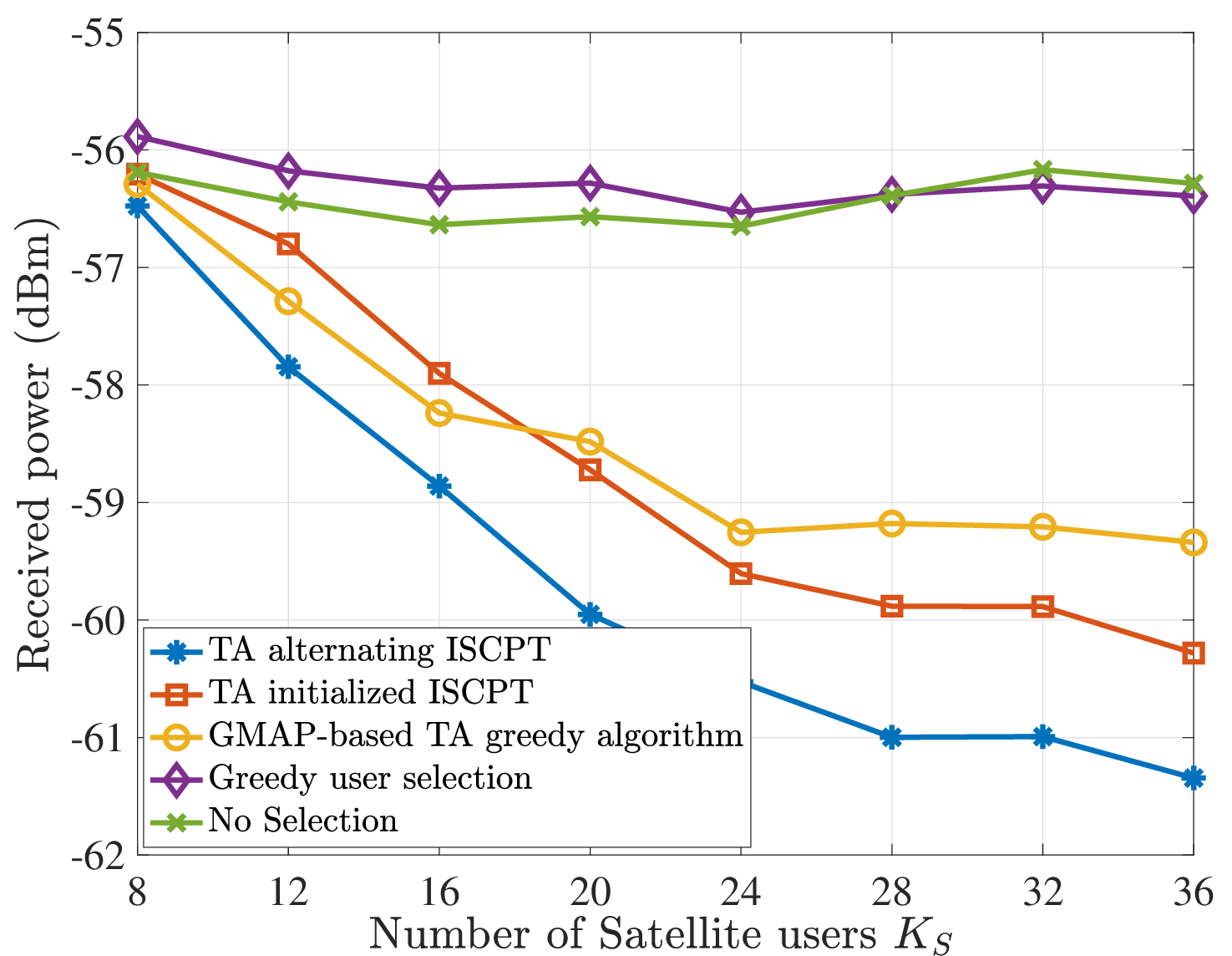}
    \caption{Power received by charging user versus the number of satellite users}
    \label{UE-3}
\end{figure}

In this section, we first present the performance of the proposed methods in terms of sum rate, sensing SINR, and received power as functions of the number of satellite users $K_S$, as shown in Figs. \ref{UE-1}, \ref{UE-2}, and \ref{UE-3}, respectively. For this evaluation, the number of satellite APs is fixed at $M=64$. Fig.  \ref{AP-2} further illustrates the sensing SINR performance under varying numbers of APs, with the number of satellite users fixed at $K_S=30$. This allows us to evaluate the impact of AP density on sensing performance.

In terms of communication performance, Fig. \ref{UE-1} shows that the proposed TA alternating ISCPT and GMAP-inspired TA greedy algorithms achieve a significant improvement in sum rate. The performance differences among them are primarily attributed to their ability to manage interference, which becomes more severe as the number of users increases. As user number increases, the sum rate of all methods tends to relatively stable, demonstrating the maximum loaded user to the corresponding method. Furthermore, the user count corresponding to the peak of each curve represents the maximum number of users that can be effectively supported by each method. The associated sum rate reflects each method’s capability to suppress interference and maintain communication quality. Specifically, for the no-selection scheme, the highest sum rate is achieved with 16 satellite users; however, the overall value is relatively low, and performance degrades rapidly as the number of users increases. This outcome highlights the significant impact of interference when no user selection strategy is employed. In contrast, the greedy user selection method yields suboptimal performance, as it tends to select only a small subset of highly orthogonal users. While this approach effectively suppresses interference, it significantly limits the number of concurrently active users, thereby reducing the pre-log factor and resulting in a lower overall sum rate.

The TA alternating optimization achieves the best sum-rate performance and maintains an increasing trend even beyond 36 users, indicating that the system can support more users with this method. Both the TA initialized optimization and greedy user selection approaches can accommodate around 20 users while keeping the performance stable as the number of users increases; however, their achievable sum rates are inferior to the other two methods. The noticeable gap between the TA alternating and TA initialized schemes highlights the significant performance improvement achieved by the alternating optimization. Finally, the GMAP-inspired TA greedy algorithm provides a satisfactory performance only slightly below that of the alternating optimization, while offering much lower computational complexity and supporting over 24 users within the system. In contrast to the greedy user selection method, the GMAP-inspired TA greedy algorithm also prefers the selection of orthogonal users, i.e., $N_{in}=1$, however, it modifies orthogonality by deactivating APs, which leads to a notable improvement as evidenced by the simulation results.

Fig. \ref{UE-2} shows the cumulative distribution function (CDF) of sensing SINR under varying numbers of satellite users. Since sensing performance is sensitive to interference from communication users, scenarios with $8$, $12$, and $16$ users are categorized as low-interference cases (solid lines), while higher user counts represent high-interference conditions (dashed lines).  
The greedy user selection and GMAP-inspired TA greedy methods achieve the highest sensing SINR, reaching approximately $15$-$20$ dB. This is primarily because they effectively suppress interference by deactivating conflicting communication users and selecting the AP with the strongest channel to the sensing target. The TA alternating ISCPT method also performs well, achieving a range of $10$-$20$ dB. However, unlike the TA greedy methods, it may not always prioritize the AP with the strongest sensing link, as it also balances communication performance. The TA initialized ISCPT optimization shows relatively poor performance overall due to the unsuitable thresholds. In scenarios with a low number of users, there is a higher probability of achieving higher sensing performance due to lower interference levels. The no-selection scheme yields the poorest performance and is highly susceptible to interference, as increasing the number of communication users significantly raises the probability of conflicts with sensing targets.

Sensing SINR is influenced by the number of satellite APs in the system. While increasing the number of APs makes the network more sparse, it can also introduce additional conflicts with the dominant links of the sensing target. This trade-off is evident in the performance of the no-selection and TA initialized ISCPT methods, as shown in Fig. \ref{AP-2}, where sensing SINR degrades with increasing the number of APs, i.e., from $16$-$128$ APs. In contrast, the greedy user selection and GMAP-inspired TA greedy algorithms achieve superior sensing SINR performance. This is attributed to their ability to exploit the increased number of APs to improve connectivity quality for the sensing target, while still avoiding interference through selective user and AP activation. The TA alternating ISCPT method achieves slightly lower performance compared to the greedy approaches, yet its CDF curves remain nearly identical across different AP counts. This consistency arises from the fact that the alternating method prioritizes joint optimization of communication performance and power transfer constraints, which may result in not always selecting the AP with the strongest sensing link.

Regarding the charging performance illustrated in Fig. \ref{UE-3}, it is evident that for the methods without AP selection, the received power remains relatively constant regardless of an increasing number of satellite users. The methods presented, including AP selection, specifically the GMAP-inspired TA greedy algorithm, TA initialized ISCPT, and TA alternating ISCPT methods, demonstrate a decrease in received power as the number of satellite users increases. The TA alternating ISCPT exhibits the lowest received power due to its requirement to optimize the communication user performance. The GMAP-inspired TA greedy method demonstrates marginally superior performance compared to the topological initialization approach. This indicates that the topological-based method tends to prioritize closing the AP to maximize the prelog factor. Nonetheless, despite a significant degradation in performance compared to the absence of AP selection, the maximum difference observed among the methods is only 5 dBm, which is relatively small. The TA alternating ISCPT method demonstrates effective performance across all three objectives and exhibits an outstanding ability to reduce user interference. The GMAP-inspired TA greedy method can provide satisfactory performance while employing significantly lower complexity. 

\section{Conclusion}
\label{sec:con}
This paper investigated the ISCPT problem in a multi-user SAGIN, in which a novel TA framework was proposed to capture the connectivity between users and satellite APs, as well as to represent interference relationships among users. The visibility and channel strength criteria were introduced to leverage the topological structure, enabling the physical interpretation of communication, sensing, and power transfer objectives within the topology graph. To jointly optimize these objectives, we reformulated the ISCPT problem as two separate yet tractable MILPs  with dedicated individual objectives, for which their alternating optimization yields better performance than each individual subproblem. Furthermore, a low-complexity greedy GMAP algorithm achieved satisfactory performance with substantially reduced computation costs. Given the demonstrated effectiveness in addressing the ISCPT problem, the proposed TA framework is expected to be one of the promising network architectures for the future SAGINs.

\vspace{-0.5cm}
\section{Appendix}
\subsection{Proof of Optimization $\Pc_1^{''}$}
\label{sec:AppA}
In \cite{yu2021active,khalilsarai2018fdd}, the transformation of the optimization objective function $\Pc_1^{'}$ has been proven and utilized extensively. Therefore, this paper omits the derivation from \eqref{Objective-P1} to \eqref{Objective-v2}, and \eqref{Matching-1-v2} to \eqref{Matching-3-v2}. To enhance the tractability of all constraints, we focus on the high SNR scheme, i.e., $\sigma^2\to \infty$. To the communication SINR, by applying \eqref{SINR-Topo} to each activated communication user, denoted as $\forall k\in\Kc_C^{'}$, enables the constraint \eqref{SINR-P1} to be reformulated as  
\begin{align}
   \frac{\sum_{m=1}^{M}v_m\hat{w}_{mk}^2}{\sum_{m=1}^{M}(\sum_{k^{'}\neq k}u_{k^{'}}v_m{\hat{w}_{mk}\hat{w}_{mk^{'}}}+v_m^s{\hat{w}_{mk}\hat{w}_{ms}})} \geq \tau_c, 
\end{align}
where $\tau_c\in(0,\infty)$. Thus, it is same with 
\begin{align}
    &\frac{\sum_{m=1}^{M}(\sum_{k^{'}\neq k}u_{k^{'}}v_m{\hat{w}_{mk}\hat{w}_{mk^{'}}}+v_m^s{\hat{w}_{mk}\hat{w}_{ms}})}{\sum_{m=1}^{M}v_m\hat{w}_{mk}^2} \leq \frac{1}{\tau_c},\\
    &\frac{\sum_{m=1}^{M}v_m\hat{w}_{mk}^2}{\sum_{m=1}^{M}(\sum_{k^{'}\in\Kc_C}u_{k^{'}}v_m{\hat{w}_{mk}\hat{w}_{mk^{'}}}+v_m^s{\hat{w}_{mk}\hat{w}_{ms}})}\geq\hat{\tau}_c,\label{SINR-Pro1}
\end{align}
where we define $\hat{\tau}_c\triangleq\frac{1}{\frac{1}{\tau_c}+1}$. Then, we have transferred the threshold $\tau_c$ to a controllable threshold $\hat{\tau}_c \in (0,1]$. Inequality \eqref{SINR-Pro1} can be rewritten into a more restrictive yet simpler constraint as: 
\begin{align}
  &\hat{w}_{mk}^2+c_1(1-v_m)+c_2(1-u_k)\geq\hat{\tau}_{c,m}\times \notag\\
&\left(\sum_{k^{'}\in\Kc_C}u_{k^{'}}{\hat{w}_{mk}\hat{w}_{mk^{'}}}+v_m^s{\hat{w}_{mk}\hat{w}_{ms}}\right),\ \forall m\in\Mc, k\in\Kc_C
\end{align}
which corresponds the constraint \eqref{SINR-v2} of $\Pc_1^{''}$. 
For the sensing SINR, by applying the sensing SINR formulation \eqref{Sensing-Topo}, constraint \eqref{Sensing-P1} can be expressed as
\begin{align}
& \frac{\sum_{m=1}^{M}v_m^s\hat{w}_{ms}^2}{\sum_{k\in\Kc_C}u_{k}\sum_{m=1}^Mv_m{\hat{w}_{mk}\hat{w}_{ms}}}\geq \tau_s. 
\end{align}
We can likewise utilize the same method to reformulate the constraint as 
\begin{align}
&\sum_{m=1}^Mv_m^s\hat{w}_{ms}^2\geq\hat{\tau}_{s,k}\left(\sum_{m=1}^Mv_m^s\hat{w}_{ms}^2+\sum_{m=1}^Mv_m\hat{w}_{mk}\hat{w}_{ms}\right), \notag\\  &\qquad\qquad\qquad\qquad\qquad\qquad\qquad\qquad\qquad\forall \ k\in\Kc_C^{'}
\end{align}
where $\hat{\tau}_{s,k}$ indicates the threshold to each communication user. This corresponds to the constraint \eqref{SINR-v2} in $\Pc_1^{''}$. The hyper-parameter $v_m^s$ specifies the choice of the sensing AP; therefore, if the $m$-th AP is selected, it must be activated. In addition, this paper focuses on selecting a single AP for the sensing target, requiring the extra two additional constraints as
\begin{align}
    &v_m^s\leq v_m\\
    &\sum_{m=1}^M [\Am]_{m,K_C+1}v_m^s = 1.
\end{align}
They correspond the constraints \eqref{Sensing-2-v2} and \eqref{Sensing-3-v2} of $\Pc_1^{''}$. To the power transfer, by applying \eqref{Power-Topo} into constraint \eqref{Power-Con}, we have
\begin{align}
\sum_{k_p\in\Kc_P}\sum_{m=1}^M\left(\sum_{k\in\Kc_C}u_{k}v_m{\hat{w}_{mk_p}\hat{w}_{mk}}+v_m^s{\hat{w}_{mk_p}\hat{w}_{ms}}\right)\geq \tau_p. 
\end{align}
In practical scenarios, the power constraint per AP has to be considered. The precoder for each AP needs to be normalized and transmitted at the specified power level. Thus, the received power of the charging user is exclusively dependent on the status of the AP. The interference suffered by sensing and communication users, resulting from the status of the AP, can be assessed through alternative constraints. Consequently, the power constraint can be reformulated as: $\sum_{k_p\in\Kc_P}\sum_{m=1}^Mv_m\hat{w}_{mk_p}\geq \hat{\tau}_p$,  
where we define $\hat{\tau}_p=\sum_{k_p\in\Kc_P}\sum_{m=1}^M\hat{w}_{mk_p}$. This indicates the proportion of power that can be transferred to the charging user via the system. 

\subsection{Proof of Optimization $\Pc_2^{''}$}
\label{sec:AppB}
The difference from optimization $\Pc_1^{''}$ lies in the objective function, which shifts from maximizing the communication sum rate to optimizing the sensing SINR. Therefore, this subsection concentrates on the reformulation of the objective. Initially, we incorporate \eqref{Sensing-Topo} into the objective function as: 
\begin{align}
    \max\ \frac{\sum_{m=1}^{M}v_m^s\hat{w}_{ms}^2}{\sum_{k\in\Kc_C}u_{k}\sum_{m=1}^Mv_m{\hat{w}_{mk}\hat{w}_{ms}}},
\end{align}
which is the same as 
\begin{align}
    \min\ \frac{\sum_{k\in\Kc_C}u_{k}\sum_{m=1}^Mv_m{\hat{w}_{mk}\hat{w}_{ms}}}{\sum_{m=1}^{M}v_m^s\hat{w}_{ms}^2}.
\end{align}
In order to transfer the objective to a tractable one, we define a hyper-parameter $t$ as 
\begin{align}
    \min_{u_k,v_m,v_m^s,t} \ &t\\
    \st&\ \frac{\sum_{k\in\Kc_C}u_{k}\sum_{m=1}^Mv_m{\hat{w}_{mk}\hat{w}_{ms}}}{\sum_{m=1}^{M}v_m^s\hat{w}_{ms}^2}\leq t.\label{Sensing-pro-1}
\end{align}
The constraint in \eqref{Sensing-pro-1} can then be written as 
\begin{align}
    \sum_{k\in\Kc_C}u_{k}\sum_{m=1}^Mv_m{\hat{w}_{mk}\hat{w}_{ms}}\leq t\sum_{m=1}^{M}v_m^s\hat{w}_{ms}^2.
\end{align}
By applying the similar transformation, this constraint is rewritten as a more restrictive one as: 
\begin{align}
    &\sum_{m=1}^Mv_m{\hat{w}_{mk}\hat{w}_{ms}}\leq t\sum_{m=1}^{M}\hat{w}_{ms}^2+c_1(1-u_k)+c_2(1-v_m^s), \notag\\
&\qquad\qquad\qquad\qquad\qquad\qquad\qquad\forall \ k\in\Kc_C,m\in\Mc
\end{align} 
and the objective function becomes: 
\begin{align}
    \min_{u_k,v_m,v_m^s,t} t,
\end{align}
which are identical to the objective in \eqref{Objective-P2} and the sensing constraint in \eqref{Sensing-1-P2}. The remaining constraints, related to the sensing SINR, communication SINR, and power transfer, are consistent with those defined in \eqref{P1}.
\vspace{-0.5cm}
\bibliography{Reference_Satellite}
\bibliographystyle{IEEEtran}

\end{document}